\documentclass[preprint,prd,amsmath,amssymb,nofootinbib,tightenlines,floatfix]{revtex4}
\usepackage[dvips]{graphics}
\usepackage{epsfig}
\usepackage{latexsym}
\usepackage{color}

\newcommand{\be}{\begin{equation}}
\newcommand{\ee}{\end{equation}}
\newcommand{\bea}{\begin{eqnarray}}
\newcommand{\eea}{\end{eqnarray}}
\newcommand{\ba}{\begin{array}}
\newcommand{\ea}{\end{array}}

\newcommand{\vect}[1]{\mathbf{#1}}

\newcommand{\abs}[1]{\left\lvert #1\right\rvert}

\newcommand{\pd}[2]{\frac{\partial #1}{\partial #2}}
\newcommand{\mcdot}{\!\cdot\!}
\newcommand{\CPV}{CP\!\!\!\!\!\!\!\!\raisebox{0pt}{\small$\diagup$}}

\begin{document}

\preprint{Caltech MAP-312}
\preprint{INT-PUB 06-03}

\title{\Large  
Yukawa and Tri-scalar Processes \\
in Electroweak Baryogenesis
}

\author{Vincenzo Cirigliano\footnote{
	Electronic address: vincenzo@caltech.edu}
}
\affiliation{California Institute of Technology, Pasadena, CA 91125}

\author{Michael J. Ramsey-Musolf\footnote{
	Electronic address: mjrm@caltech.edu}}
\affiliation{California Institute of Technology, Pasadena, CA 91125}

\author{Sean Tulin\footnote{
	Electronic address: tulin@caltech.edu}
}
\affiliation{California Institute of Technology, Pasadena, CA 91125}

\author{Christopher Lee\footnote{
	Electronic address: clee@phys.washington.edu}
}
\affiliation{Institute for Nuclear Theory, University of Washington,
Seattle, WA 98195}

\begin{abstract}

We derive the contributions to the quantum transport equations for
electroweak baryogenesis due to decays and inverse decays induced by
tri-scalar and Yukawa interactions. 
In the Minimal Supersymmetric Standard Model (MSSM), these
contributions give rise to couplings between Higgs and fermion
supermultiplet densities, thereby communicating the effects of CP-violation in the Higgs sector to the baryon sector. We show that the decay and inverse decay-induced contributions that arise
at zeroth order in the strong coupling,
$\alpha_s$,  can be substantially larger than the ${\cal
O}(\alpha_s)$ terms that are generated by scattering processes and that are usually assumed
to dominate.  We revisit the often-used approximation of fast Yukawa-induced
processes and show that for realistic parameter choices it is not
justified.  We solve the resulting quantum transport equations
numerically with special attention to the impact of Yukawa rates and
study the dependence of the baryon-to-entropy ratio $Y_B$ on MSSM
parameters.

\end{abstract}

\maketitle


\section{Introduction} 

The origin of the baryon asymmetry of the universe (BAU) remains an
open question for particle physics, nuclear physics, and
cosmology. Although the size of the BAU cannot be explained within the
framework of the Standard Model (SM), there exist a variety of SM
extensions that may allow for successful baryogenesis. Scenarios in
which the BAU is produced at the electroweak phase transition are
particularly attractive since they can be tested with laboratory
experiments. To the extent that the masses of the particles
responsible for baryogenesis are not too different from the weak
scale, their dynamics can be studied using a combination of collider
experiments, precision electroweak measurements, and CP-violation
studies.

In order to carry out robust tests of electroweak baryogenesis (EWB),
it is necessary to delineate systematically the quantitative
relationship between EWB and experimentally accessible
observables. The motivation for doing so has been heightened by the
prospect of significant new experimental information in the near
term. Studies at the Large Hadron Collider will search for the
existence of new particles at the TeV scale. At the same time, a new
generation of searches for permanent electric dipole moments of the
electron, neutron, and neutral atoms will look for the effects of
\lq\lq new" CP-violation with several orders of magnitude better
sensitivity than given by current experimental limits (see, {\em e.g.}, Refs.~\cite{Erler:2004cx,Pospelov:2005pr}   and references therein). Should either
the LHC or EDM searches discover evidence for new physics at the
electroweak scale, then precision studies at both the International
Linear Collider and low-energy facilities should provide detailed
information about the structure of the new physics. To the extent that
the theoretical treatment of EWB is on sufficiently firm ground, these
experimental efforts may either confirm or rule out this paradigm for
the BAU.

The basic physical picture of EWB was developed over a decade
ago~\cite{Cohen:1994ss,HN,Joyce:1994fu,Joyce:1994zn} 
(see~\cite{Trodden:1998ym} for a review). The elements
include a first-order electroweak phase transition, in which bubbles
of broken electroweak symmetry expand and fill the universe as it
cools through the transition temperature. CP- and C-violating
interactions between fields in the plasma at the phase boundary create a net chiral
charge that is injected into the region of unbroken electroweak
symmetry, driving the weak sphaleron processes that create non-zero
baryon number density, $n_B$. The expanding bubbles then capture
the non-zero $n_B$ in the region of broken electroweak symmetry,
where weak sphaleron processes are highly suppressed and unable to
affect $n_B$ appreciably. It is crucial that the first-order phase
transition be sufficiently strong
in order to preclude \lq\lq wash out" of non-zero baryon number.

The earliest analyses based on this picture employed conventional
transport theory to compute the production, diffusion, and relaxation
of chiral charge at the phase boundary.  Several groups
have subsequently endeavored to put these computations on a more sophisticated
footing by using non-equilibrium quantum field theory techniques.  As
first pointed out by Riotto~\cite{riotto0,riotto}, only a
non-equilibrium field-theoretic formulation can properly account for
the quantum nature of CP violation as well as the decoherence effects
due to the presence of spacetime-varying background fields and the
thermal bath of particles at the phase
boundary.  Using these methods Riotto~\cite{riotto} observed that
conventional treatments may overlook significant enhancement of the
CP-violating source terms in the transport equations associated with
memory effects in the plasma. The presence of such enhancements could
relax the requirements on new CP-violation needed for successful EWB,
thereby allowing for consistency between the BAU and considerably
smaller EDMs than previously thought.  This work was followed by the
authors of Refs.~\cite{Carena:2000id,Carena:2002ss}, who adopted a
similar approach to that of Ref.~\cite{riotto} in computing the
CP-violating source terms while carrying out a more comprehensive
phenomenology. The analyses of both groups were performed within the
Minimal Supersymmetric Standard Model (MSSM). 
The non-equilibrium approach has also been pursued in 
Refs.~\cite{Konstandin:2004gy,Konstandin:2003dx,Konstandin:2005cd}.

Recently, we investigated the CP-conserving terms as well as the
CP-violating sources in the transport equations using non-equilibrium
field theory methods~\cite{us}. We found that there exists a hierarchy
of physical scales associated with the electroweak phase transition
dynamics that allows one to derive the transport equations from the
Closed Time Path Schwinger-Dyson equations using a systematic
expansion in scale ratios. Again in the MSSM, we computed the
CP-violating sources and leading CP-conserving chiral relaxation terms
associated with interactions of fermion and Higgs superfields with the
spacetime-varying Higgs vacuum expectation values (vevs). Our results
for the sources were consistent with those obtained in previous
work~\cite{Carena:1997gx,riotto,Carena:2000id,Carena:2002ss,Balazs:2004ae},
but we also found that enhancements in the relaxation rates could
mitigate the effect of enhancements in the sources.

A number of other contributions to the transport equations remain to
be analyzed using non-equilibrium methods. Here, we focus on terms
that link the dynamics of the quark supermultiplets with those of the
Higgs scalars and their Higgsino superpartners. Importantly, these
terms are responsible for communicating CP-violating effects in the
Higgs supermultiplet densities to the quark supermultiplet densities,
thereby allowing CP-violating interactions in the Higgs sector to
contribute to baryogenesis.  In the MSSM, the requirement of a strong
first-order phase transition (shown in \cite{Laine} to occur in the presence of a light right-handed stop) and constraints from precision
electroweak data (requiring the left-handed stop to be heavy \cite{Carena:2000id}) imply that it is the CP-violating interactions of the
Higgs superfields -- rather than those directly involving the squarks
-- that drives baryogenesis via this coupling between the two
sectors. In extensions of the MSSM, such as the NMSSM or U(1)$^\prime$
models, the phenomenological requirements that preclude large effects
from CP-violation in the squark sector can be relaxed \cite{nonminimal}, and in this
case it is important to know the relative importance of Higgs sector
CP-violation. In either case, an analysis of the dynamics whereby the
baryon and Higgs sectors communicate is an important component of a
systematic, quantitative treatment of EWB.

Before providing the details of our study, we summarize the primary
results, using the transport equation for the Higgs $+$ Higgsino
densities for illustration:
\be
\label{eq:Heq}
\partial^\mu H_\mu = - \Gamma_H\frac{H}{k_H}
-\Gamma_Y\biggl(\frac{Q}{k_Q} - \frac{T}{k_T} + \frac{H}{k_H}\biggr) -
{\tilde\Gamma}_Y\biggl(\frac{B}{k_B} - \frac{Q}{k_Q} +
\frac{H}{k_H}\biggr)+ \bar\Gamma_Y\frac{h}{k_h} + S_{\tilde
H}^{\CPV} \ \ \ .  
\ee 
Here, $H$ and $h$ are number densities associated various combinations
of the up- and down-type Higgs supermultiplets in the MSSM (defined
below); $H_\mu$ is the corresponding vector current density; $Q$ and
($B$,$T$) are the number densities of particles in the third
generation left- and right-handed quark supermultiplets, respectively;
the $k_{H,h,Q,T,B}$ are statistical weights; $S_{\tilde H}^{\CPV}$
is a CP-violating source; and $\Gamma_H$, $\Gamma_Y$,
${\tilde\Gamma}_Y$, and $\bar\Gamma_Y$ are transport coefficients.

Physically, the presence of $S_{\tilde H}^{\CPV}$ results from an
imbalance between the rates for particle and antiparticle 
scattering off the bubble wall, favoring the generation of non-vanishing
supermultiplet densities $H$ and $h$. In contrast, the terms
proportional to $\Gamma_H$ and $\bar\Gamma_Y$ cause these densities to
relax to zero. The terms containing $\Gamma_Y$ and ${\tilde\Gamma}_Y$
favor chemical equilibrium between Higgs superfield densities and
those associated with quark supermultiplets. To the extent that the
rates $\Gamma_Y$ and ${\tilde\Gamma}_Y$ are fast compared to the rate
of relaxation, any non-vanishing Higgs supermultiplet density quickly
induces non-vanishing densities for quark supermultiplets, thereby
facilitating EWB. Understanding the microscopic dynamics of this
competition between CP-violating sources, relaxation terms, and
Higgs-baryon sector couplings is essential to achieving a quantitative
description of EWB.

In previous work, we computed $\Gamma_H$ and $S_{\tilde H}^{\CPV}$
using the Closed Time Path Schwinger-Dyson equations and considering
the lowest-order couplings between superfields and the spacetime
varying Higgs vevs.  Here, we focus on the terms proportional to
$\Gamma_Y$, $\bar\Gamma_Y$, $\tilde\Gamma_Y$ that are generated by
$Hqq$ Yukawa couplings, the corresponding supersymmetric interactions,
and the SUSY-breaking triscalar couplings\footnote{In previous work,
only the contributions to terms of this type generated by Standard
Model Yukawa interactions were considered, leading to the use of the
subscript \lq\lq $Y$".}.  We make several observations regarding these
terms:
\begin{itemize}

\item[(i)] In previous treatments, $\Gamma_Y$ and ${\bar\Gamma}_Y$
were estimated from scattering processes such as $t_R+g\to t_L+H_u^0$,
making them proportional to one power of the strong coupling,
$\alpha_s$.  We find, however,
that there exist contributions to $\Gamma_Y$ occurring at zeroth order
in $\alpha_s$ that are generated by decay and inverse decay processes
such as $t_R + t_L \leftrightarrow H_u^0$. To the extent that the
three-body processes are kinematically allowed, their contribution to
$\Gamma_Y$ can be considerably larger than those generated by
scattering. We also show that $|{\bar\Gamma}_Y/\Gamma_Y|$ is typically
$< 1/2$ for MSSM parameters consistent with precision electroweak data
and the existence of a strong first order phase transition.  (The authors of Ref.~\cite{Carena:2000id}
argued that $|{\bar\Gamma}_Y| = |\Gamma_Y|$.) We solve
the transport equations numerically and find that inclusion of the
three-body contributions affects the baryon-to-entropy ratio $Y_B$ at
the $10-20\%$ level for realistic choices of the MSSM parameters. We
provide a detailed analysis of the dependence of $Y_B$ on $\Gamma_Y$
and the MSSM parameters that determine it.

\item[(ii)] In most of the early studies of EWB in the MSSM, it was
assumed that the rate $\Gamma_Y$ of Yukawa-induced processes is
``fast'' compared to all other relevant time-scales, implying that the
Yukawa-induced transfer of non-zero Higgs/Higgsino density to
non-vanishing chiral charge density is more efficient than
relaxation. This assumption has motivated an expansion in powers of $1/\Gamma_Y$. We show that there exist corrections to the Higgs density at linear order in this expansion that have not been included in previous treatments. After including these terms, we find that the expansion itself breaks down -- even for the enhanced values of $\Gamma_Y$
that result from inclusion of the three-body contributions --  due to the presence of chirality-changing
processes in the bubble wall whose rates $\Gamma_H$ and $\Gamma_M$ can
be larger than $\Gamma_Y$.  We study numerically the impact of keeping
a finite $\Gamma_Y$: we find that the corrections to the
$\Gamma_Y \to \infty$ limit of $Y_B$ range between $20 \%$ and $100
\%$, depending on the values of the other rates.

\item[(iii)] The terms containing $\Gamma_Y$ and $\bar\Gamma_Y$ have
been included in the earlier studies of 
Refs.~\cite{Cline:2000kb,Cline:2000nw,Carena:2000id,Carena:2002ss}
\footnote{In the notation of Ref.~\cite{Carena:2000id},
${\bar\Gamma_Y}=-\rho\Gamma_Y$.}, whereas the one involving
$\tilde\Gamma_Y$ is new. In the MSSM, one often assumes that the
triscalar coupling involving the down-type Higgs scalars, the doublet
scalars ${\tilde Q}$, and the right handed scalars $\tilde b$ is
proportional to the bottom Yukawa coupling, $y_b$. For $\tan\beta \sim
{\cal O}(1)$, one has $y_b/y_t \ll 1$ and the impact of the
$\tilde\Gamma_Y$ term is relatively minor. For scenarios with large
$\tan\beta$, however, $y_b$ need not be small compared to $y_t$. In
this case the transport coefficient ${\tilde\Gamma}_Y$ and other terms
(not shown) that couple to the $B$ supermultiplet need not be
suppressed, and the coupled set of transport equations must be
augmented to include dynamical $b$-quarks and their
superpartners. Although in the present study we do not consider this
large $\tan\beta$ scenario, we provide the general formulas that allow
one compute ${\tilde\Gamma}_Y$.

\end{itemize}

In the remainder of the paper, we discuss our detailed analysis of the
$\Gamma_Y$-type terms that lead to these observations. In Section
\ref{sec:formalism}, we consider these terms for generic Yukawa and
tri-scalar interactions and analyze their dependence on the relevant
mass parameters. In Section \ref{sec:mssm} we specify to the MSSM,
including detailed analytic and numerical studies. Here, we include
contributions from both SM particles and their superpartners (in
contrast to previous anlayses that included only SM scattering terms),
and note that the superpartner contributions tend to increase the
magnitude of $\Gamma_Y$. In Section \ref{sec:solving} we solve the
coupled transport equations to obtain the baryon-to-entropy ratio, and
show why one would not expect an expansion in $1/\Gamma_Y$ to yield a
reasonable approximation to the exact solution. We summarize this work
in Section \ref{sec:conclusions}. Various technical points are
discussed in the Appendices.

\section{Three-body source terms: building blocks} 
\label{sec:formalism}

Our approach for deriving the source terms in the quantum transport
equations is based on the Closed Time Path Schwinger-Dyson
equations. An extensive discussion of this framework is
given in our earlier work \cite{us}. Here, we give a brief summary of
our method and use it to derive the source terms generated by
supersymmetric Yukawa and SUSY-breaking tri-scalar interactions to
leading order in the loop expansion.

\subsection{Formalism and method}

Ordinary quantum field theory is not appropriate for treating the
microscopic dynamics of the electroweak phase transition (EWPT), since
the non-adiabatic evolution of states and the presence of degeneracies
in the spectrum break the zero-temperature, equilibrium relation
between the in- and out-states.  The non-adiabaticity arises because
particle interactions occur against a spacetime-varying background
field (the Higgs vevs), while thermal effects associated with
non-zero temperature introduce degeneracies in the spectrum. The
impact of non-adiabaticity and degeneracies on quantum evolution can
be treated systematically using the Closed Time Path (CTP) 
formalism~\cite{CTP}.
In this formulation the time arguments of all fields and composite
operators lie on a path ${\cal P}$ that consists of a positive branch
${\cal P}_{+}$ from $-\infty$ to $+\infty$ and a negative branch
${\cal P}_{-}$ running back from $+\infty$ to $-\infty$.  Fields whose
arguments lie on ${\cal P}_{+}$ precede those on ${\cal P}_{-}$ along
the path ${\cal P}$. Moreover, those lying on ${\cal P}_{+}$ are time-ordered while those on ${\cal P}_{-}$ are anti-time-ordered.

With this prescription the standard time-ordering operator $T$ is
replaced by the path-ordering operator $T_{\cal P}$ and the
perturbative expansion is formally identical to the equilibrium case. In applying Wick's theorem, however, one must allow for contractions involving all possible combinations of fields taken from either 
${\cal P}_{+}$ and ${\cal P}_{-}$, leading to a generalized Green's function that accounts for path-ordering. Specifically, the bosonic and fermionic Green functions are given by 
\begin{eqnarray}
G(x,y) &=& \langle  T_{\cal P}\, \left[ \phi (x) \phi^\dagger (y) 
\right] \rangle 
\label{eq:gfb}
\\
S(x,y) &=& \langle  T_{\cal P}\, \left[ \psi (x) \overline{\psi} (y) 
\right] \rangle  
\label{eq:gff}
\end{eqnarray}
where $\langle ... \rangle $ denotes an average over the physical state
of the system, which may be described by an appropriate density
matrix.  In practical applications it is convenient to use ordinary
time arguments, in terms of which each of Eq.~(\ref{eq:gfb}) and
(\ref{eq:gff}) represents four Green functions and decomposes in 
various components. To establish the notation we recall here 
explicitly the bosonic Green functions: 
\begin{eqnarray}
G^{++}(x,y) &\equiv&  G^{t}(x,y) =
 \langle  T \, \left[ \phi (x) \phi^\dagger (y) \right] \rangle 
\\
G^{+-}(x,y) &\equiv&  G^<(x,y) =
 \langle \phi^\dagger (y) \phi(x) \rangle 
\\
G^{-+}(x,y) &\equiv&  G^>(x,y) =
\langle \phi (x) \phi^\dagger (y) \rangle 
\\
G^{--}(x,y) &\equiv&  G^{\bar{t}}(x,y) =
 \langle \bar{T} \, \left[ \phi (x) \phi^\dagger (y) \right] \rangle \ \ \ ,
\end{eqnarray}
where the superscripts \lq\lq $a$" and \lq\lq  $b$" in $G^{ab}(x,y)$ indicate the branch ${\cal P}_{\pm}$
on which the time components of $x$ and $y$ lie, respectively and where ${\bar T}$ is the anti-time-ordering operator.

\begin{figure}[!t]
\centering
\epsfig{figure=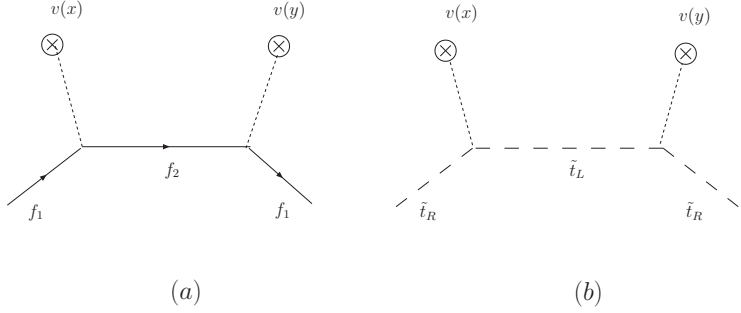,width=10cm}
\caption{
Leading contributions to the self-energies $\Sigma_{B,F}$
generated by scattering from Higgs vevs.
\label{fig:vevs}
}
\end{figure}

The equations governing the spacetime dependence of number densities
of a given bosonic or fermionic species can be derived from the
Schwinger-Dyson equations for the generalized Green's functions
$G(x,y)$ and $S(x,y)$ and have the following form
\cite{kadanoff-baym,riotto}:
\begin{equation}
\begin{split}
\pd{n_B}{X_0} (X) +{\mbox{\boldmath$\nabla$}}\mcdot\vect{j}_B(X) 
= \int d^3 z\int_{-\infty}^{X_0} dz_0\
\Bigl[ \Sigma_B^>(X,z) G^<(z,X)&-G^>(X,z)\Sigma_B^<(z,X)\\
+G^<(X,z) \Sigma_B^>(z,X) &- \Sigma_B^<(X,z) G^>(z,X)\Bigr]\,.
\label{eq:scalar1}
\end{split}
\end{equation}
\begin{equation}
\begin{split}
\pd{n_F}{X_0} (X) + {\mbox{\boldmath$\nabla$}}\mcdot\vect{j}_F(X) =  
-\int d^3 z\int_{-\infty}^{X_0} dz_0\
{\rm Tr}\Bigl[ \Sigma_F^>(X,z) S^<(z,X)&-S^>(X,z)\Sigma_F^<(z,X)\\
+S^<(X,z) \Sigma_F^>(z,X) &- \Sigma_F^<(X,z) S^>(z,X)\Bigr]\, .
\label{eq:fermion1}
\end{split}
\end{equation}
The RHS involves a causal time integral over the system's history and
is expressed in terms of the Green functions (\ref{eq:gfb}),
(\ref{eq:gff}) and self energies $\Sigma_{B,F}$ that encode all the
information about particle interactions.  This feature allows for a
consistent treatment of both CP-violating terms ``sourcing'' a given
particle density as well as CP-conserving interactions that tend to
transfer this density to other species or cause it to relax away.
Previously, the leading CP-violating contributions to $\Sigma_{B,F}$
generated by scattering from the Higgs vevs (see
Fig.~\ref{fig:vevs}) were computed in
Refs.~\cite{riotto,Carena:2002ss,Carena:2000id,Balazs:2004ae,us} while
the corresponding CP-conserving relaxation terms generated by the same
processes were derived in Ref.~\cite{us}.  Here, we extend these
analyses to include the three-body source terms that arise from Yukawa
and tri-scalar interactions at one-loop order (see
Fig. \ref{fig:fig1}).

In general, the Green's functions (\ref{eq:gfb}), (\ref{eq:gff}) are
dynamical objects that can be obtained by solving the transport
equations (\ref{eq:scalar1}) and (\ref{eq:fermion1}). However, the
hierarchy of time and energy scales present during the electroweak 
phase transition allow for simplifications in
treating the transport equations \cite{us}. The time scales are a decoherence
time, $\tau_{\rm d}$, associated with the departure from adiabatic
evolution; a \lq\lq plasma" time, $\tau_{\rm p}$, associated with
mixing between degenerate states in the finite temperature spectrum; and
the \lq\lq intrinsic" quasiparticle evolution time, $\tau_{\rm int}$,
associated with time evolution of a state of definite energy. In terms
of physical parameters associated with the plasma, one has $\tau_{\rm
d}\sim 1/(v_w k_{\rm eff})$, $\tau_{\rm p}\sim 1/\Gamma_p$, and
$\tau_{\rm int}\sim 1/\omega$, where $v_w$ is the velocity of
expansion of the bubble wall; $k_{\rm eff}$ is an effective wave
number that in general depends on the quasiparticle wave number and
wall thickness, $L_w$; $\Gamma_p$ is the thermal width of the
quasiparticle; and $\omega$ is the quasiparticle frequency that
depends on both the particle momentum and thermal mass. 
For the EWPT,
one has that $\varepsilon_{\rm d}=\tau_{\rm int}/\tau_{\rm d} \ll 1$;
$\varepsilon_{\rm p}=\tau_{\rm int}/\tau_{\rm p} \ll 1$; and
$\varepsilon_{\rm d}/\varepsilon_{\rm p} = \tau_{\rm p}/\tau_{\rm d}
\ll 1$. 
In addition, the small densities present at the EWPT imply a
hierarchy of energy scales: $\varepsilon_\mu = \mu/T \ll 1$, where
$\mu$ refers to the chemical potential of any particle species. The
existence of these hierarchies allows for a number of simplifying
approximations in solving the transport equations:

\begin{itemize}
\item[(i)] Use of the quasiparticle ansatz for the $G_{i}(x,y)$. This
relies on $\varepsilon_{\rm p} \ll 1$, that is, that  the damping rates $\Gamma_p$ that broaden the
spectrum of excitations are typically suppressed when compared with
the excitation frequencies ($\Gamma_{i}/\omega_i \ll 1$).

\item[(ii)] Working near kinetic and chemical equilibrium.  This
approximation relies on $\tau_{\rm p}/\tau_{\rm d} \ll 1$, that is, the
plasma interactions among quasiparticles are fast
compared to the decoherence time, thereby leading to approximate,
local equilibrium among quasiparticle species. Consequently, one may
approximate quasiparticle distribution functions appearing in the
Green functions by their equilibrium forms and track quasiparticle
densities with local chemical potentials.  The error engendered by
doing so is ${\cal O}(\varepsilon_{\rm d}/\varepsilon_{\rm p})$ and is, thus, negligible.
\end{itemize} 

Motivated by these considerations we evaluate the source terms on
the RHS of the transport equations (\ref{eq:scalar1}) and
(\ref{eq:fermion1}) using the free-particle form of the Green
functions. For example, for the boson Green functions, we have
\begin{eqnarray}
G_i^> (x,y) &=& \int \frac{d^4 k}{(2 \pi)^4} e^{-ik\cdot(x-y)}\Bigl[1 + f_B(k_0,\mu_i) \Bigr]
\ \rho_i(k_0,{\bf k})
\label{eq:gf>}
 \\
G_i^< (x,y) &=& \int \frac{d^4 k}{(2 \pi)^4} e^{-ik\cdot(x-y)} f_B(k_0,\mu_i) 
\ \rho_i(k_0,{\bf k})
\label{eq:gf<}
\end{eqnarray}
with spectral functions $\rho_i(k_0,{\bf k})= \pi/\omega_{\bf k}
\left[ \delta(k^0 - \omega_{\bf k}) - \delta(k^0 + \omega_{\bf
k})\right]$ ($\omega_{\bf k} = \sqrt{{\bf k}^2 + m^2}$) that can be
appropriately modified to take into account collision-broadening and
thermal masses, and distribution functions close to the
equilibrium form
\begin{equation}
f_B(k_0,\mu_i) = n_B(k_0,\mu_i) + {\cal O}(\varepsilon_{\rm d}/\varepsilon_{\rm p}) \ \ \ ,
\label{eq:distr}
\end{equation}
where $ n_B (k_0,\mu_i) = 1/[e^{(k_0 - \mu_i)/T} - 1] $ and $\mu_i$ is
a local chemical potential .

Upon expanding the source terms to lowest non-trivial order in
the $\varepsilon_{{\rm d},{\rm p},\mu}$ and relating
current and chemical potential to local densities through 
\begin{equation}
\label{eq:nrelation}
{\bf j}_i (X) = - D_i \, \nabla n_i (X)
\qquad \qquad n_i (X) = \frac{T^2}{6} \, k_i (m_i/T) \, \mu_i (X) ,
\end{equation}
where $k_i(m_i/T)$ is a statistical factor (see, \emph{e.g.} \cite{us}), 
we obtain the quantum transport equations\footnote{ The quantum
transport equations (\ref{eq:bde}) are sometimes referred to as quantum
Boltzmann or diffusion equations.}
\begin{equation}
\dot{n}_i  - D_i \nabla^2 n_i = S_i[ \{ n_j \}] \ . 
\label{eq:bde}
\end{equation}
In Eq.~(\ref{eq:bde}) both CP-violating effects and relaxation rates
are encoded in the quantum mechanical sources $S_i[\{n_j \}]$.

\subsection{Results for generic tri-scalar and  supersymmetric Yukawa 
interactions}

\begin{figure}[!t]
\centering
\epsfig{figure=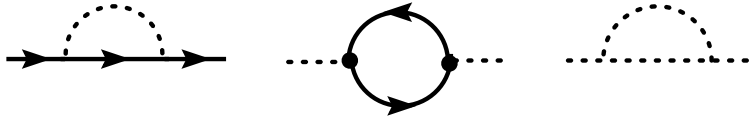,width=10cm}
\caption{
Self energies for scalar and fermion fields 
induced by the Yukawa and triscalar interaction lagrangians 
of Eqs.~(\ref{eq:lagYs}) and (\ref{eq:lagYf}). 
\label{fig:fig1}
}
\end{figure}

Let us consider now the generic three-scalar interaction,
\begin{equation}
\mathcal{L}_{\text{int}} = \lambda_s A_s  
\phi_L  \phi_R^* \phi_H + \text{h.c.}\, ,
\label{eq:lagYs}
\end{equation}
where $\lambda_s$ is a dimensionless coupling and $A_s$ is a mass
scale\footnote{In the MSSM, $\lambda_s$ is the Yukawa coupling and $A_s$ is either the $\mu$-parameter or the soft, tri-scalar coupling.}.  This interaction generates contributions to the self-energy
appearing on the RHS of Eq.~(\ref{eq:scalar1}) through the one-loop diagram
depicted in Fig.~\ref{fig:fig1}(c).  As an example, we give the
self-energy for the complex scalar $\phi_R$, 
\begin{equation}
\Sigma_{R}^{>,<} (x,y) = - |\lambda_s A_s |^2 \, G_L^{>,<} (x,y)
\, G_H^{>,<} (x,y) \ , 
\label{eq:sescalar}
\end{equation}
Importantly, the RHS of Eq.~(\ref{eq:sescalar}) is manifestly
independent of possible CP-violating phases appearing in the coupling
$\lambda_s A_s$ and therefore does not contribute to the CP-violating
source.  We obtain similar results for the self-energies of $\phi_L$ and
$\phi_H$.  This situation contrasts with that for the Higgs vev
scattering contributions derived from Fig.~\ref{fig:vevs}, where
interference terms involving the up- and down-type Higgs vevs at the
different vertices contain CP-violating phase effects.

Inserting Eq.~(\ref{eq:sescalar}) into Eq.~(\ref{eq:scalar1}), using
the Green Functions of
Eqs.~(\ref{eq:gf>},\ref{eq:gf<},\ref{eq:distr}), and expanding to
first order in $\varepsilon_\mu$ and zeroth order in $\varepsilon_{\rm
d,p}$ (setting the thermal widths to zero), we obtain the
leading, three-body contribution to $S_R$ on the RHS of
Eq.~(\ref{eq:bde}).  We find that the three-body sources for the
particle number densities of the complex scalars $\phi_L$, $\phi_R$
and $\phi_H$ are related to each other and are given by
\begin{equation}
\begin{split}
S_R(X) &= -S_L(X) = -S_H(X) \\
&= - \left( \mu_R-\mu_L-\mu_H \right) (X) \ \abs{\lambda_s}^2  \
\mathcal{I}_B (A_s \, ; \,   m_R,m_L,m_H)  \ , 
\end{split}
\label{eq:sourceS}
\end{equation}
in terms of the function
\begin{equation}
\begin{split}
& \mathcal{I}_B (A_s \, ; \,   m_R,m_L,m_H) = - \frac{\abs{A_s}^2}{16\pi^3 T}
 \,
\int_{m_R}^\infty d\omega_R \int_{\omega_L^-}^{\omega_L^+} d\omega_L  \\
& \ \ \ \times \biggl\{n_B(\omega_R)\bigl[1+n_B(\omega_L)\bigr] 
n_B(\omega_L-\omega_R)\bigl[\theta(m_R-m_L-m_H) - \theta(m_L-m_R-m_H)\bigr] \\
& \qquad  -n_B(\omega_R)n_B(\omega_L)\bigl[
1+n_B(\omega_L+\omega_R)\bigr]\theta(m_H-m_R-m_L)\biggr\},
\end{split}
\label{eq:IB}
\end{equation}
with integration limits given by:
%
%
\begin{equation}
\begin{split}
\omega_L^\pm = \frac{1}{2m_R^2}\Bigl\{&\omega_R\abs{m_R^2 + m_L^2 - m_H^2}  \\
&\pm \bigl[(\omega_R^2 - m_R^2)\bigl(m_R^2 - (m_L + m_H)^2\bigr)\bigl(m_R^2-(m_L - m_H)^2\bigr)\bigr]^{1/2}\Bigl\}.
\end{split}
\end{equation}

The presence of mass thresholds and combinations of Bose distributions
in Eq.~(\ref{eq:IB}) makes clear its interpretation in terms of
physical processes in the plasma: decay $R \to L + H$ and all possible
emission/absorption channels that are kinematically allowed.  It is
straightforward to integrate over $\omega_L$ and obtain a
representation of the source in terms of one-dimensional integrals.
We give this formula in Appendix~\ref{app:app1}.  Finally, we note
that $S_{R,L,H}$ are of first order in the $\varepsilon$ counting
discussed above, whereas the leading CP-violating sources and
CP-conserving relaxation terms generated from the tree-level graphs of
Fig.~\ref{fig:vevs} are ${\cal O}(\varepsilon_{\rm d}\varepsilon_{\rm
p})$ and ${\cal O}(\varepsilon_{\rm p}\varepsilon_{\mu})$,
respectively. Nonetheless, they can be similar in magnitude to
$S_{R,L,H}$ since the latter contain additional phase-space
suppression factors $\sim 16\pi$ associated with the absorptive part
of one-loop graphs.

We now consider contributions from a generic Yukawa interaction
\begin{equation}
\mathcal{L}_{\text{int}} = \lambda_f 
(\phi\bar\psi_1 P_L\psi_2 + \phi^*\bar\psi_2 P_R\psi_1)
\label{eq:lagYf}
\end{equation}
that generates contributions to both scalar and fermionic self-energies on
the RHS of Eqs.~(\ref{eq:scalar1}) and (\ref{eq:fermion1}) through the
diagrams depicted in Fig.~\ref{fig:fig1}(a),(b).
The resulting  source for particle number densities associated 
with the complex scalar $\phi$ and Dirac fermions $\psi_1$ and 
$\psi_2$ are related to each other and read:
\begin{equation}
\begin{split}
S_\phi(X) &= -S_{\psi_1}(X) = S_{\psi_2}(X) \\
&= - \left(\mu_\phi-\mu_1+\mu_2 \right)(X) \ \abs{\lambda_f}^2 \ 
 \mathcal{I}_{F} (m_1,m_2,m_\phi) 
\ , 
\end{split}
\label{eq:sourceF}
\end{equation}
where 
\begin{equation}
\begin{split}
& \mathcal{I}_F (m_1,m_2,m_\phi) = \frac{1}{16\pi^3 T} \, 
\left(m_1^2 + m_2^2 - m_\phi^2 \right) \, 
\int_{m_1}^\infty d\omega_1 \int_{\omega_\phi^-}^{\omega_\phi^+}d\omega_\phi \\
&\quad\times\biggl\{n_B(\omega_\phi)\bigl[1-n_F(\omega_1)\bigr]n_F(\omega_1-\omega_\phi)\bigl[\theta(m_1-m_2-m_\phi) - \theta(m_\phi - m_1 - m_2)\bigr] \\
&\qquad + n_B(\omega_\phi)n_F(\omega_1)\bigl[1-n_F(\omega_1+\omega_\phi)\bigr]\theta(m_2-m_1-m_\phi)\biggr\}
\end{split}
\label{eq:IF}
\end{equation}
with integration limits on $\omega_\phi$ given by:
%
\begin{equation}
\begin{split}
\omega_\phi^\pm = \frac{1}{2m_1^2}\Bigl\{&\omega_1\abs{m_\phi^2 + m_1^2 - m_2^2} \\
& \pm \bigl[(\omega_1^2 - m_1^2)\bigl(m_1^2-(m_2 + m_\phi)^2\bigr)\bigl(m_1^2-(m_2 - m_\phi)^2\bigr)\bigr]^{1/2}\Bigl\}.
\end{split}
\end{equation}
As in the bosonic case, Eq.~(\ref{eq:IF}) has a direct interpretation
in terms of decay, emission and absorption of $\phi$, $\psi_1$ and
$\psi_2$ in the plasma.  The integration over $\omega_\phi$ is
straightforward and we report its result in Appendix \ref{app:app1}.

\begin{figure}[!t]
\centering
\begin{picture}(300,170)  
\put(0,60){\makebox(50,50){\epsfig{figure=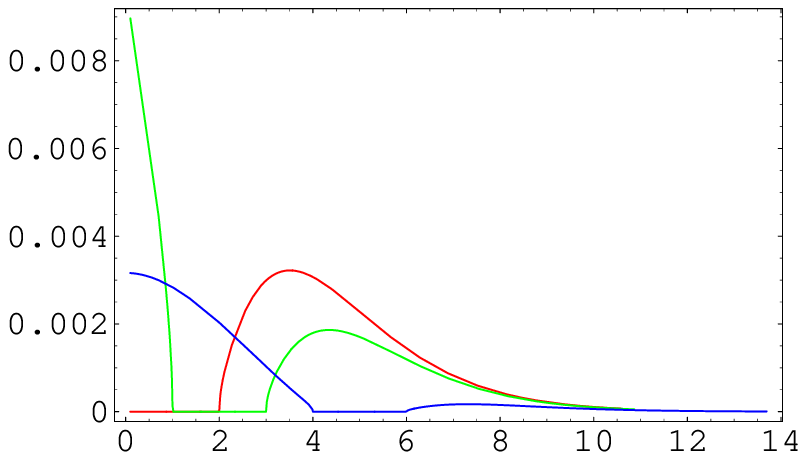,width=7.5cm}}}
\put(240,60){\makebox(50,50){\epsfig{figure=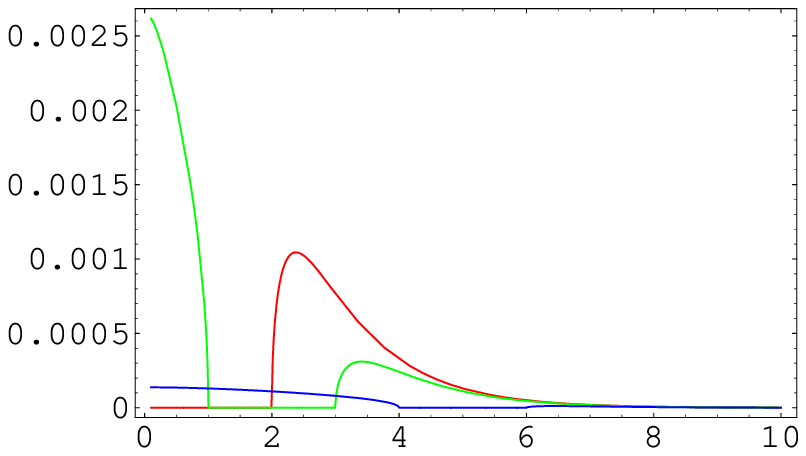,width=7.5cm}}}
\put(110,10){{$\displaystyle m_\phi/T$ }}
\put(350,10){{$\displaystyle m_H/T$ }}
\put(-45,125){{\tiny $\frac{m_2}{T} =2$}}
\put(0,75){{\tiny $\frac{m_2}{T} =1$}}
\put(-45,70){{\tiny $\frac{m_2}{T} =5$}}
\put(200,125){{\tiny $\frac{m_R}{T} =2$}}
\put(245,75){{\tiny $\frac{m_R}{T} =1$}}
\put(200,50){{\tiny $\frac{m_R}{T} =5$}}
\put(-90,155){{ $ {\mathcal I}_F/T^3 $}}
\put(150,155){{ ${\mathcal I}_B/T^3 $ }}
\end{picture}
\caption{
Left panel: $\mathcal{I}_F/T^3$ as a function of $m_\phi/T$ for 
$m_1/T = 1$ and $m_2/T = 1,2,5$. 
Right panel: 
$\mathcal{I}_B/T^3$ as a function of $m_H/T$ for 
$A_s/T=1$, $m_L/T=1$ and $m_R/T=1,2,5$. 
\label{fig:fig2}
}
\end{figure}

Eqs.~(\ref{eq:sourceS}-\ref{eq:IB}) and (\ref{eq:sourceF}-\ref{eq:IF})
are central new results of this paper and represent the building
blocks out of which we can construct the three-body physical sources
in the MSSM (Section \ref{sec:mssm}).  In order to identify the
dominant contributions to the MSSM sources, where many individual
building blocks contribute, it is instructive to characterize the
behavior of $\mathcal{I}_{B,F}$ as function of the masses of the
interacting particles.  The main features are:

\begin{itemize}

\item[i)] Symmetry properties under exchanges $m_L \leftrightarrow m_R 
\leftrightarrow m_H$ and 
$m_1 \leftrightarrow m_2$:
$$ \mathcal{I}_B (A_s
 \, ; \, m_R,m_L,m_H) = \mathcal{I}_B (A_s \, ; \, m_L,m_R,m_H)
\qquad  
\mathcal{I}_F (m_1,m_2,m_\phi) = \mathcal{I}_F (m_2,m_1,m_\phi) $$

\item[ii)] There are threshold effects which can be read off via the
explicit $\theta$-functions.  In order for the rate to be non-zero,
the mass arguments have to be such that at least one of the two body
decays $a \to b+c$ is kinematically allowed.

\item[iii)] $ \mathcal{I}_B (A_s \, ; \, m_R,m_L,m_H)$ and $\mathcal{I}_F
(m_1,m_2,m_\phi)$ are largest when the three masses are such that the
largest mass is slightly greater than the sum of the two smaller ones
(just above threshold).

\item[iv)] 
$ \mathcal{I}_B (A_s \, ; \, m_R,m_L,m_H)$ and $\mathcal{I}_F
(m_1,m_2,m_\phi) $ become vanishingly small as any of the masses
becomes much larger than the temperature. This reflects Boltzmann
suppression of the thermally averaged rate.
Moreover, $\mathcal{I}_F (m_1,m_2,m_\phi)$ vanishes as all the masses
become much smaller than the temperature.

\end{itemize}

The above properties are illustrated in Fig.~\ref{fig:fig2}, where we
plot $\mathcal{I}_F/T^3$ as a function of $m_\phi/T$ for representative
choices of $m_1$ and $m_2$ (left panel) and similarly 
$\mathcal{I}_B/T^3$ as a function of $m_H/T$ for representative choices of
$m_L$, $m_R$, and $A_s$ (right panel).

\section{Three-body source terms in the MSSM}
\label{sec:mssm}

The results of Sec.~\ref{sec:formalism} allow us to calculate the
sources for quark, squark, Higgs, and Higgsino particle densities
generated by the supersymmetric Yukawa and SUSY-breaking triscalar
interactions in the MSSM. We focus on those involving the third-generation quark supermultiplets whose interactions generally depend
on the large Yukawa coupling $y_t$. As noted in the Introduction,
these interactions dominate for $\tan\beta\sim{\cal O}(1)$, whereas
interactions proportional to $y_b$ can be important for large
$\tan\beta$. While the results in the previous section would allow us
to compute these $y_b$ effects -- such as the transport coefficient
${\tilde \Gamma}_Y$ appearing in Eq. (\ref{eq:Heq}) -- including them
would lead to a more complex set of coupled transport equations. For
simplicity, we focus here on the smaller $\tan\beta$ case with
$a_f\propto y_f$ -- wherein interactions involving $y_t$ dominate --
and defer a more general treatment to a future study.

\subsection{Interactions in the MSSM}

The terms in the MSSM superpotential generating interactions proportional to $y_t$ are:
\begin{equation}
W = y_t Q_3 H_u \bar t_R + \mu H_u H_d,
\end{equation}
where the weak doublets are defined $Q_3 = (t_L, b_L)$, $H_u = (H_u^+, H_u^0)$, and $H_d = (H_d^0, H_d^-)$
In addition the soft SUSY-breaking Lagrangian contains the terms:
\begin{equation}
\mathcal{L}_{\text{soft}} = - a_t \tilde Q_3 H_u \tilde t_R^* + \text{h.c.}
\end{equation}
In the minimal supergravity (mSUGRA) scenario for SUSY breaking, the
$a$-parameters are proportional to the Yukawa couplings,
e.g. $a_t = y_t A_t$ for some mass parameter $A_t$. Thus this part of
$\mathcal{L}_{\text{soft}}$ also generates contributions to the top
three-body source that are proportional to $y_t$.

From both the supersymmetric and soft SUSY-breaking sectors, we obtain  the tri-scalar interactions
\begin{equation}
\label{eq:LYscalar}
\mathcal{L}_{\text{scalar}}^Y = - y_t \tilde t_R^*\tilde t_L(A_t H_u^0 + \mu^*H_d^{0*}) + y_t\tilde t_R^* \tilde b_L (A_t H_u^+ - \mu^* H_d^{-*}) + \text{h.c.},
\end{equation}
and the supersymmetric Yukawa interactions 
\begin{equation}
\begin{split}
\mathcal{L}^Y_{\text{fermion}} &= y_t(-H_u^0 t_R^\dag t_L + H_u^+ t_R^\dag b_L) 
+ y_t(-\tilde t_R t_L^\dag \tilde H_u^{0\dag} + \tilde t_R b_L^\dag \tilde H_u^{+\dag}) \\
&\quad + y_t(-\tilde t_L t_R^\dag \tilde H_u^0 + \tilde b_L t_R^\dag \tilde H_u^+) + \text{h.c.}
\end{split}
\end{equation}
In order to write this Lagrangian in the form appearing in
Eq.~(\ref{eq:lagYf}), we combine the two-component Higgsino spinors
into four-component Dirac spinors, which is sensible in the unbroken
electroweak phase where the mass terms for Higgsinos are simply:
\begin{equation}
\mathcal{L}^{\tilde H}_{\text{mass}} = -\mu\tilde
H_u^+\tilde H_d^- + \mu\tilde H_u^0\tilde H_d^0 +
\text{h.c.}
\end{equation}
First rotating the fields ${\tilde H}_u^{0,+}\rightarrow
e^{-i\phi_\mu}{\tilde H}_u^{0,+}$ to remove the complex phase from
$\mu$, we define the Dirac spinors
\begin{equation}
\label{eq:DiracPsiH}
\Psi_{\tilde H^+} = \begin{pmatrix} \tilde H_u^+ \\ \tilde
H_d^{-\dag} \end{pmatrix} \qquad \Psi_{\tilde H^0} =
\begin{pmatrix} -\tilde H_u^0 \\ \tilde H_d^{0\dag}
\end{pmatrix},
\end{equation}
which have Dirac mass $\abs{\mu}$.  We define chemical potentials
$\mu_{\tilde H^+}, \mu_{\tilde H^0}$ corresponding to the
vector charge densities $\bar\Psi\gamma^0\Psi$ for these Dirac
fields. In terms of these fields, the Yukawa interaction terms are:
\begin{equation}
\label{eq:LYfermion}
\begin{split}
\mathcal{L}^Y_{\text{fermion}} = &y_t(-H_u^0 \bar t_R P_L t_L + H_u^+
\bar t_R P_L b_L) \\ &+ y_t e^{i\phi_\mu}(\tilde t_R \bar t_L P_R
\Psi_{\tilde H^0}^C + \tilde t_R\bar b_L P_R \Psi_{\tilde
H^+}^C) \\ &+ y_t e^{-i\phi_\mu}(\tilde t_L\bar t_R
P_L\Psi_{\tilde H^0} + \tilde b_L\bar t_R P_L\Psi_{\tilde
H^+}) + \text{h.c.},
\end{split}
\end{equation}
making use also of the charge-conjugated fields:
\begin{equation}
\Psi_{\tilde H^\pm}^C = \begin{pmatrix} \tilde H_d^- \\
\tilde H_u^{+\dag} \end{pmatrix} \qquad \Psi_{\tilde H^0}^C =
\begin{pmatrix} \tilde H_d^0 \\ -\tilde H_u^{0\dag}
\end{pmatrix},
\end{equation}
where $\Psi^C = C\bar\Psi^T$, with $C = i\gamma^2\gamma^0$.

\subsection{Source Terms in the MSSM}

Having identified the relevant interactions in the MSSM Lagrangian
proportional to $y_t$ in Eqs.~(\ref{eq:LYscalar}) and
(\ref{eq:LYfermion}), we can write the sources for the densities of
the particles appearing in these interactions using the general
results of Eqs.~(\ref{eq:sourceS},\ref{eq:sourceF}).

To be concrete, let us focus on the right-handed top squark and quark
densities. Similar formulas will hold for the left-handed squarks and
quarks, and the Higgs and Higgsinos. The source for the right-handed
top squark number density $n_{\tilde t_R}$ is:
\begin{equation}
\label{eq:tRsource}
\begin{split}
S^Y_{\tilde t_R}(X) = -N_C y_t^2\Bigl[&(\mu_{\tilde t_R} - \mu_{\tilde
t_L} - \mu_{H_u^0})\mathcal{I}_B(A_t ; m_{\tilde t_R}, m_{\tilde t_L},
m_{H_u^0}) \\ +&(\mu_{\tilde t_R} - \mu_{\tilde b_L} - \mu_{H_u^+})
\mathcal{I}_B(A_t ; m_{\tilde t_R}, m_{\tilde b_L}, m_{H_u^+}) \\
+&(\mu_{\tilde t_R} - \mu_{\tilde t_L} + \mu_{H_d^0})\mathcal{I}_B(\mu
; m_{\tilde t_R}, m_{\tilde t_L} , m_{H_d^0}) \\ +&(\mu_{\tilde t_R} -
\mu_{\tilde b_L} + \mu_{H_d^-})\mathcal{I}_B(\mu ; m_{\tilde t_R},
m_{\tilde b_L}, m_{H_d^-}) \\ +&(\mu_{\tilde t_R} - \mu_{t_L} -
\mu_{\tilde H^0})\mathcal{I}_F(m_{\tilde H^+}, m_{t_L}, m_{\tilde
t_R}) \\ + &(\mu_{\tilde t_R} - \mu_{b_L} - \mu_{\tilde
H^+})\mathcal{I}_F(m_{\tilde H^0}, m_{b_L}, m_{\tilde t_R})\Bigr],
\end{split} 
\end{equation}
and for the quark density $n_{t_R}$:
\begin{equation}
\label{eq:stRsource}
\begin{split}
S^Y_{t_R}(X) = - N_C y_t^2\Bigl[&(\mu_{t_R} - \mu_{t_L} - \mu_{H_u^0})
\mathcal{I}_F(m_{t_R},m_{t_L},m_{H_u^0}) \\ + &(\mu_{t_R} - \mu_{b_L}
- \mu_{H_u^+}) \mathcal{I}_F(m_{t_R}, m_{b_L}, m_{H_u^+}) \\ +
&(\mu_{t_R} - \mu_{\tilde t_L} - \mu_{\tilde H^0})
\mathcal{I}_F(m_{t_R}, m_{\tilde H^0}, m_{\tilde t_L}) \\ +
&(\mu_{t_R} - \mu_{\tilde b_L} - \mu_{\tilde H^+})
\mathcal{I}_F(m_{t_R}, m_{\tilde H^+}, m_{\tilde b_L})\Bigr]
\end{split}
\end{equation}

The various chemical potentials appearing in the source can be related
by making the assumption, first introduced in Ref.~\cite{HN}, 
of fast gauge and gaugino interactions and zero density of gauge
bosons or gauginos ($\mu_V = \mu_{\tilde V} = 0$). In this case, pairs
of superpartner densities are in chemical equilibrium, as are members
of the same gauge multiplet. Thus,
\begin{subequations}
\label{eq:equalmus}
\begin{align}
\mu_{t_R} &= \mu_{\tilde t_R} \equiv \mu_T \\
\mu_{t_L} &= \mu_{\tilde t_L} = \mu_{b_L} = \mu_{\tilde b_L} \equiv \mu_Q \\
\mu_{H_u^0} &= \mu_{H_u^+} \equiv \mu_{H_u} \\
\mu_{H_d^0} &= \mu_{H_d^-} \equiv \mu_{H_d} \\
\mu_{\tilde H^+} &= \mu_{\tilde H^0} \equiv \mu_{\tilde H} \label{eq:equalHiggsinomus}
\end{align}
\end{subequations}
Relating the scalar Higgs chemical potentials $\mu_{H_{u,d}}$ to the
Higgsino chemical potential $\mu_{\tilde H}$ is somewhat more
subtle and we refer to Appendix~\ref{app:app2} for a derivation. 
Defining the combinations, 
\begin{eqnarray}
\label{eq:vectorscalarrelation_def}
\mu_H & \equiv & \frac{1}{2}\left(\mu_{H_u} - \mu_{H_d}\right) 
\\
\mu_{h} &\equiv& \frac{1}{2}\left(\mu_{H_u} + \mu_{H_d}\right),
\end{eqnarray}
the supergauge equilibrium condition reads:
\begin{equation}
\label{eq:higgseq}
\mu_H  = \mu_{\tilde H} ~ . 
\end{equation}

As noted in previous work, the assumption of supergauge equilibrium --
together with the relations (\ref{eq:equalmus}, \ref{eq:higgseq}) --
suggest combining the various particle densities in equilibrium with
one another into:
\begin{subequations}
\begin{align}
T &\equiv n_{t_R} + n_{\tilde t_R} \\
Q &\equiv n_{t_L} + n_{b_L} + n_{\tilde t_L} + n_{\tilde b_L} \\
H &\equiv n_{H_u^+} + n_{H_u^0} - n_{H_d^-} - n_{H_d^0} + n_{\tilde H^+} 
+ n_{\tilde H^0} \\
h &\equiv n_{H_u^+} + n_{H_u^0} + n_{H_d^-} + n_{H_d^0} 
\end{align}
\end{subequations}

Adding together the top and stop sources in
Eq.~(\ref{eq:tRsource},\ref{eq:stRsource}) and using the relations
among chemical potentials
(\ref{eq:equalmus}, \ref{eq:vectorscalarrelation_def}, \ref{eq:higgseq})
leads to the Yukawa source for the density $T$ reported in
Eq.~(\ref{eq:Tsource}) of Appendix~\ref{app:app2}.  Finally, by noting
that the masses of weak doublet partners are the same (see
Eq.~(\ref{eq:massrelations})) and converting the chemical potentials
to densities using Eq.~(\ref{eq:nrelation}) we obtain
\begin{equation}
S_T^Y(X) = -\Gamma_Y\biggl(\frac{T}{k_T} - \frac{Q}{k_Q} - 
\frac{H}{k_H}\biggr) - \bar\Gamma_Y \frac{h}{k_h},
\end{equation}
where
\begin{subequations}
\label{eq:GammaY}
\begin{align}
\begin{split}
\Gamma_Y = \frac{12 N_C y_t^2}{T^2}\Bigl[&\mathcal{I}_B(A_t ;
m_{\tilde t_R}, m_{\tilde Q}, m_{H_u}) + \mathcal{I}_B(\mu ;
m_{\tilde t_R}, m_{\tilde Q}, m_{H_d}) \\ + &\mathcal{I}_F(\mu ,
m_Q, m_{\tilde t_R}) + \mathcal{I}_F(m_{t_R}, m_Q, m_{H_u}) +
\mathcal{I}_F(m_{t_R}, \mu, m_{\tilde Q})\Bigr]
\end{split} \\
\bar\Gamma_Y = \frac{12 N_C y_t^2}{T^2}\Bigl[&\mathcal{I}_B(\mu;
m_{\tilde t_R}, m_{\tilde Q}, m_{H_d}) - \mathcal{I}_B(A_t;
m_{\tilde t_R}, m_{\tilde Q}, m_{H_u}) -
\mathcal{I}_F(m_{t_R},m_Q, m_{H_u})\Bigr]
\end{align}
\end{subequations}
Similar formulas hold for the sources $S_{Q,H,h}^Y$.

\subsection{Transport Equations and study of Yukawa rates}

Incorporating the Yukawa contributions to the sources into the full
set of transport equations for the densities $T,Q,H$ derived in
Ref.~\cite{us}, we obtain
\begin{subequations}
\label{eq:QTEs}
\begin{align}
\begin{split}
\partial^\mu T_\mu = &- \Gamma_M^-\biggl(\frac{T}{k_T} -
\frac{Q}{k_Q}\biggr) + \Gamma_M^+\biggl(\frac{T}{k_T} +
\frac{Q}{k_Q}\biggr) + S_{\tilde t}^{\CPV} \\
&-\Gamma_Y\biggl(\frac{T}{k_T} - \frac{Q}{k_Q} - \frac{H}{k_H}\biggr)
- \bar\Gamma_Y\frac{h}{k_h} + \Gamma_{ss}\biggl(\frac{2Q}{k_Q} -
\frac{T}{k_T} + \frac{9(Q+T)}{k_B}\biggr)
\end{split} \\
\begin{split}
\partial^\mu Q_\mu = &- \Gamma_M^-\biggl(\frac{Q}{k_Q} -
\frac{T}{k_T}\biggr) - \Gamma_M^+\biggl(\frac{T}{k_T} +
\frac{Q}{k_Q}\biggr) - S_{\tilde t}^{\CPV} \\
&-\Gamma_Y\biggl(\frac{Q}{k_Q} - \frac{T}{k_T} + \frac{H}{k_H}\biggr)
+ \bar\Gamma_Y\frac{h}{k_h} - 2\Gamma_{ss}\biggl(\frac{2Q}{k_Q} -
\frac{T}{k_T} + \frac{9(Q+T)}{k_B}\biggr)
\end{split} \\
\partial^\mu H_\mu = &- \Gamma_H\frac{H}{k_H}
-\Gamma_Y\biggl(\frac{Q}{k_Q} - \frac{T}{k_T} + \frac{H}{k_H}\biggr) +
\bar\Gamma_Y\frac{h}{k_h} + S_{\tilde H}^{\CPV}
\end{align}
\end{subequations}
In addition, there should be one more equation for $\partial^\mu
h_\mu$, but we have left for future work the calculation of the
$CP$-violating contribution, $S_{\tilde h}^{\CPV}$ to its source, as
well as its relaxation coefficient $\Gamma_h$.

The structure of the transport equations (\ref{eq:QTEs}) is similar to
that of the equations derived in the treatment of
Ref.~\cite{HN,Carena:2000id,Carena:2002ss}. However, use of the CTP
framework leads to a number of new features that we highlight:

\begin{itemize}

\item[i)] The appearance of new combinations of densities that do not
arise in earlier treatments -- such as those involving $\Gamma_M^+$ --
follows from a systematic treatment of the CTP Schwinger-Dyson
equations.

\item[ii)] The Yukawa rates $\Gamma_Y$ and ${\bar\Gamma}_Y$ arise at
lower-order in $\alpha_s$ than the corresponding terms in previous
treatments.  As indicated in the Introduction, these rates were
calculated to ${\cal O}(\alpha_s)$ from scattering processes such as
$t_R+g\to t_L+H_u^0$ and only the contributions from Standard Model
particles were included.  We have included here the ${\cal
O}(\alpha_s^0)$ contributions generated by decays and inverse decays
within the plasma, which -- when not vanishing due to threshold effects --
can be of comparable size or larger than the ${\cal O}(\alpha_s)$
scattering terms.  This can be appreciated by comparing the behavior of
$\Gamma_Y$ from decays (Eq.~(\ref{eq:GammaY})) and from 
scattering  (see Ref.~\cite{Joyce:1994fu,Joyce:1994zn}):
\begin{eqnarray}
\Gamma_Y^{\rm  decays}  &=&    \frac{1}{8 \pi^3} \,   y_t^2 \, 
 \frac{\tilde{M}^2}{T} \times \mathcal{O}(1) \ ,  \\
\Gamma_Y^{\rm  scattering}  &=&    \frac{\zeta_3}{6 \pi^3} \, 
 g_s^2  y_t^2 T   \   \log \left( 
\frac{8 T^2}{m_q ^2(T)} \right) \simeq     \frac{\zeta_3}{6 \pi^3} 
 \, y_t^2 T   \times \mathcal{O}(1)  ~ , 
\end{eqnarray}
where $\tilde{M}$ is a typical (thermal) mass of the order of the
electroweak scale (could be a soft SUSY breaking mass term),
$m_q(T)$ is the quark thermal mass, and $\zeta_3 = 1.202$.

\item[iii)] Because we have included both SM particle and superpartner
contributions, $\Gamma_Y$ and ${\bar\Gamma}_Y$ display a non-trivial
dependence on the MSSM parameters.  Similar observations have been
made about the CP-violating sources \cite{riotto,Carena:2000id,us} and
leading chiral relaxation terms \cite{us}, for which the possibility
of resonant enhancements have been observed.  We note that the
enhancements of the CP-violating sources and chiral relaxation are
generally not accompanied by resonant enhancements of the $\Gamma_Y$
and ${\bar\Gamma}_Y$ terms, thereby leading to a more subtle
competition between the effects of CP-violation, chiral relaxation,
and density transfer.
\end{itemize}

\begin{figure}[!t]
\centering
\epsfig{figure=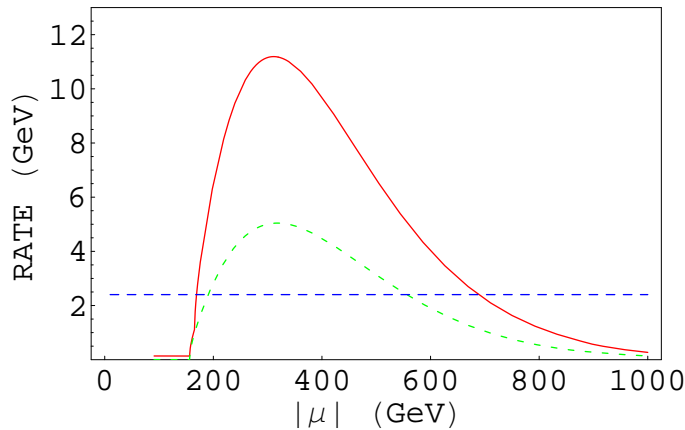,width=10cm}
\caption{ $\Gamma_Y$ (solid red line), $\bar{\Gamma}_Y$ (dashed green
line), and $\Gamma_Y^{\rm scattering}$ (dashed straight blue line) in
units of GeV as a function of $\mu$ (GeV), for 
$T=100$ GeV and SUSY mass parameters as described in the text.  
In large regions of parameter space we find $\Gamma_Y^{\rm decay} >
\Gamma_Y^{\rm scattering}$.
\label{fig:gammay}
}
\end{figure}

A quantitative illustration of the above points ii) and iii) is given
in Fig.~\ref{fig:gammay}, where we plot $\Gamma_Y$ and
$\bar{\Gamma}_Y$ versus the MSSM parameter $|\mu|$ for $T=100$ GeV.
In the numerical evaluation we include thermal masses as calculated
in~\cite{Enqvist:1997ff} and, for illustrative purposes, we use the
weak-scale SUSY parameters given in Table \ref{tab:susy}
consistent with electroweak symmetry breaking, a strongly first-order
electroweak phase transition and electroweak precision tests.  The
non-trivial $\mu$ dependence displayed by $\Gamma_Y^{\rm decay} $ is
due to threshold effects in the functions ${\mathcal I}_{B,F}$.  The
dashed straight line in Fig.~\ref{fig:gammay} represents
$\Gamma_Y^{\rm scattering}$.  In large regions of parameter space we
find $\Gamma_Y^{\rm decay} > \Gamma_Y^{\rm scattering}$.

We conclude this Section by noting that for typical values of SUSY
parameters, the chiral relaxation rates $\Gamma_{M,H}$ (active only in
the broken electroweak phase) are of comparable size or larger than
$\Gamma_Y$.  All of these rates, in turn, are much larger than the
diffusion rates $v_w^2/D_i$, which for typical values of the diffusion
constants~\cite{Joyce:1994fu,Joyce:1994zn} and wall velocity vary in
the range $10^{-3}-10^{-2}$ GeV.  We  discuss the consequence of
this when solving the diffusion equations in the following Section.

\section{Solving the Transport Equations and Phenomenology}
\label{sec:solving}

The baryon asymmetry is seeded by the density of left-handed weak
isodoublets $n_L = 5 Q + 4 T$~\cite{Cohen:1994ss,HN}, which we obtain by solving
the transport equations (\ref{eq:QTEs}). In this section we study the
impact of $\Gamma_Y$ on the solution of the system (\ref{eq:QTEs}) and
on the overall baryon-to-entropy ratio $Y_B \equiv n_B/s$.

Before entering the details of our analysis, let us shortly recall the
basic notation (see \cite{us} and references therein) and describe the
input MSSM parameters which will be used in the subsequent
numerical explorations. The baryon-to-entropy ratio can be expressed
as an integral of $n_L = 5 Q + 4 T$ in the unbroken phase:
\be
\label{eq:rhob3}
Y_B = - \frac{n_F \Gamma_{\rm ws}}{2 s} \, \frac{1}{D_q 
\lambda_+}  \int_{-\infty}^{0}  \ 
n_L (x) 
\,  e^{- \lambda_{-}\,  x } 
dx  \ . 
\ee
where $\Gamma_{\rm ws}$ is the weak sphaleron rate $\Gamma_{\rm ws} =
6 \kappa \alpha_w^5 T$ (with $\kappa \simeq 20$~\cite{wsrate}), $n_F$
is the number of fermion families, $D_q$ is the quark diffusion
constant, $v_w$ is the wall velocity and
\begin{eqnarray}
\lambda_{\pm} &=&\frac{1}{2 D_q} \,  
\left( v_w \pm \sqrt{v_w^2 + 4 D_q {\cal R}}  \right) 
\nonumber \\ 
{\cal R} &=& \Gamma_{\rm ws}\, \left[
\frac{9}{4} \, \left(1 + \frac{ n_{\rm squark}}{6}\right)^{-1} + \frac{3}{2}
\right] ~ ,
\end{eqnarray} 
where $n_{\text{squark}}$ is the number of flavors of light squarks.
Isolating the dependence on the CP-violating phases $\phi_\mu$ and
$\phi_A$, $Y_B$ is conveniently parameterized as follows~\cite{us}:
\be
\label{eq:pheno1} 
Y_{B} = F_1 \, \sin \phi_\mu \ + \ F_2 \, \sin \left( \phi_\mu +
\phi_A \right)  \ , 
\ee
in terms of $F_{1}$ (arising from the Higgsino source) and $F_{2}$
(arising from the squark source).  

In all the plots reported in this section, we adopt for the weak-scale
SUSY parameters the {\it reference values} reported in
Table~\ref{tab:susy}, which are consistent with a strongly first-order
electroweak phase transition and the constraints from precision
electroweak physics as well as direct searches.
\begin{table}[t]
\begin{center}
\begin{tabular}{|c|} 
\hline 
$\tan \beta = 10$ \\
$M_{\tilde{t}_R} = 0$  \\
$M_{\tilde{Q}} = 1$ TeV \\
$M_2 = A_t = 200$ GeV  \\
$m_{H_u}^2 = - (100 \, {\rm GeV})^2$ \\
$m_{H_d}^2 = 0$ \\
$m_A = 150$ GeV \\ 
$100 \ {\rm GeV} < |\mu| < 400 \ {\rm GeV}$  \\
\hline
\end{tabular}
\end{center}
\caption{\label{tab:susy}
Reference values of weak-scale SUSY parameters. 
}
\end{table}
Note that a CP-odd Higgs mass $m_A = 150$ GeV translates into
$\Delta \beta \sim 0.015$~\cite{bubble}.  From the reference values of
Table~\ref{tab:susy} one can derive typical values for the bubble wall
velocity and thickness, for which we use $v_w = 0.05$
~\cite{John:2000zq} and $L_w = 25/T$ ~\cite{bubble}. With this choice
of parameters one has $F_2 \sim 10^{-3} F_1$.

We now discuss in greater detail the role of Yukawa-induced
rates on the transport equations. 

\subsection{Revisiting the approximation of fast $\mathbf{\Gamma_Y}$:
need for numerical solution}

Starting with the work~\cite{HN},  the conventional
practice has been to solve the system of transport equations (\ref{eq:QTEs})
under the assumption that the rate $\Gamma_Y$ of Yukawa-induced
processes~\footnote{ As well as the rate $\Gamma_{ss}$ of strong
sphaleron processes.} is ``fast'' compared to all other relevant
time-scales, thereby ensuring a chemical equilibrium condition among
$H$, $Q$, and $T$.  Doing so allows one to obtain analytic expressions
for $Y_B$.  The assumption of fast Yukawa interactions is  well
justified in the unbroken phase ahead of the advancing bubble wall, where a
particle may diffuse for a period characterized by the 
the inverse of the diffusion rate $\Gamma_{\rm diff} =
v_w^2/D$ before the bubble wall catches it. In order for Yukawa processes to be effective in this region, they must act quickly on the time scale $\Gamma_{\rm diff}^{-1}$, and one, indeed, finds that 
$\Gamma_Y \gg \Gamma_{\rm diff}$ for
typical values of the diffusion
constants~\cite{Joyce:1994fu,Joyce:1994zn} and wall
velocity.
In the broken electroweak
phase, however, $\Gamma_{\rm diff}^{-1}$ is no longer the only 
relevant time scale. In addition, Yukawa processes must compete with scattering from the spacetime-varying Higgs vevs that leads  to relaxation of chiral charge and Higgs supermultiplet  densities. Importantly, 
the corresponding rates ($\Gamma_M$ and $\Gamma_H$, respectively)
are as large as or larger than
$\Gamma_Y$ -- even after including the ${\cal O}(\alpha_s^0)$
contributions to $\Gamma_Y$.  As a result,  the interplay of
these competing processes within the bubble wall is  significant, and 
imposing the condition of $\Gamma_Y$-induced chemical equilibrium is not justified.\footnote{In addition, the authors of Ref.~\cite{Cline:2000kb} noted that the condition $\Gamma_Y\rightarrow\infty$ causes a parametric suppression of the Higgs source $h$, while for realistic parameter choices, the suppression factor turns out to $\mathcal{O}(1)$.}
%

To make this key  point more explicit,  we have solved 
the transport equations in powers of $1/\Gamma_{Y,ss}$ and analyzed the magnitude of the corrections
to the $\Gamma_{Y,ss} \to \infty$ limit. Explicit details are given in Appendix~\ref{app:app3}, where 
we point out that the most important correction was missed
in previous analytic approaches to this problem -- namely 
the correction to the $H$ density induced by an effective shift 
in the source $S_{\tilde H}^{\CPV}$.
The analysis of Appendix~\ref{app:app3} implies that fractional
corrections to the baryon asymmetry to first order in $1/\Gamma_Y$
read
\begin{equation}
\label{eq:solve9text}
\frac{\delta Y_B}{Y_B}\sim
\left(\frac{\Gamma_H}{\Gamma_Y}\right)
\frac{\sqrt{r_\Gamma} \, \Gamma_M^- \, L_w}{\sqrt{\bar{D} (\Gamma_M^- + 
\Gamma_H)}} 
\end{equation}
where $r_\Gamma \sim 0.07$~.  Substituting the earlier estimates of $\Gamma_H$ and $\Gamma_M^-$~\cite{HN} into this expression, we
find $\delta Y_B/Y_B \sim 0.1$ -- indeed a small correction.  However,
when using $\Gamma_H$, $\Gamma_M^-$ and $\Gamma_Y$ as calculated in
Ref.~\cite{us} and the present work within the CTP framework, we find
much larger corrections: $\delta Y_B/Y_B \sim \mathcal{O}(1)$.  This difference
is  due primarily to the larger values of $\Gamma_H$ and $\Gamma_M$
obtained in our framework~\cite{us} (even off resonance) compared to
previous calculations~\cite{Joyce:1994fu,Joyce:1994zn,HN}.

\begin{figure}[!t]
\centering
\epsfig{figure=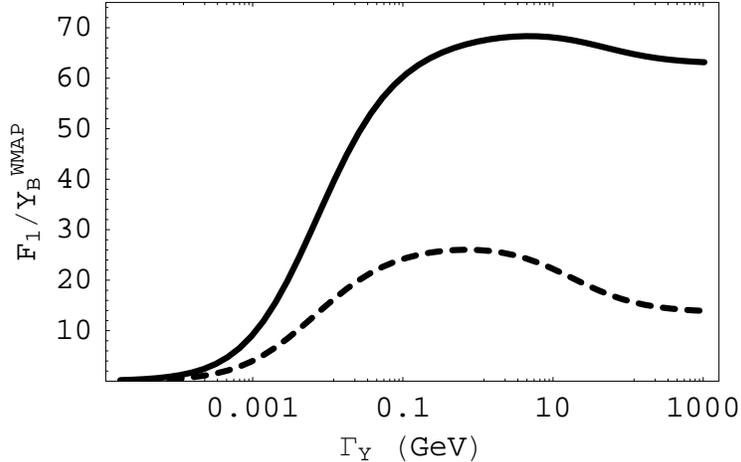,width=10cm}
\caption{ We plot here the ratio $F_1/Y_B^{\rm WMAP}$ versus
$\Gamma_Y$ for two values of the SUSY $\mu$ parameter: $|\mu| = 200$
GeV (solid line) and $|\mu| = 250$ GeV (dashed line), corresponding to
on-resonance and off-resonance baryogenesis, respectively. All other
parameters are fixed at the reference values of 
Table~\ref{tab:susy}. We use the central value $Y_B^{\rm WMAP}=9.2
\times 10^{-11}$~\cite{wmap}. 
\label{fig:F1vsGammaY}
}
\end{figure}

The above considerations imply that in order to avoid $\mathcal{O}(1)$
uncertainties in the calculation of $Y_B$, one requires a full numerical
solution of the system (\ref{eq:QTEs}).  In order to quantify the
effect, we plot  in Fig.~\ref{fig:F1vsGammaY}  the ratio $F_1/Y_B^{\rm
WMAP}$ 
versus $\Gamma_Y$
for two values of the SUSY $\mu$ parameter: $|\mu| = 200$ GeV (solid
line) and $|\mu| = 250$ GeV (dashed line), corresponding to
on-resonance and off-resonance baryogenesis, respectively. All other
parameters are fixed as in Table~\ref{tab:susy}. 
Typical values of $\Gamma_Y$ lie in the range $5-10$ GeV (see
Fig.~\ref{fig:gammay}).  The curves in Fig.~\ref{fig:F1vsGammaY}  illustrate
two key points of the Yukawa-induced dynamics:
\begin{itemize}
\item[(i)] Efficient chargino/neutralino-mediated  baryogenesis occurs for
$\Gamma_Y \gg v_w^2/D_h \sim 0.0025$ GeV, as the Higgs supermultiplet
density $H$ injected in the unbroken phase is efficiently converted
into LH top-quark density (fueling sphaleron processes) before the bubble
catches up. Inclusion of the $\mathcal{O}(\alpha_s^0)$ terms in $\Gamma_Y$
affects $Y_B$ at the $10-20 \%$ level, as one is already in the 
plateau region in Fig.~\ref{fig:F1vsGammaY}.

\item[(ii)] As $\Gamma_Y$ increases (keeping all other rates fixed)
the baryon asymmetry reaches a maximum and then starts decreasing
towards its asymptotic value.  This behavior can be 
understood qualitatively as follows. In the non-resonant case (dashed line),  as $\Gamma_Y$ increases, Yukawa induced processes start to complete with $\Gamma_{H}$ {\em inside} the bubble
wall, thereby transferring $H$ density to $Q,T$ densities.  The latter subsequently 
relax away due to $\Gamma_M^-$ processes or diffuse very
inefficiently into the unbroken phase.  This effect is less pronounced in the resonant 
case (solid line), where $Y_B$ first grows as $\Gamma_Y$ becomes more efficient compared to diffusion ahead of the bubble wall, but  then saturates due to the presence of resonantly-enhanced Higgs supermultiplet relaxation within the plasma.

\end{itemize}

Summarizing, the main message emerging from
Fig.~\ref{fig:F1vsGammaY} is the following: keeping $\Gamma_Y$ finite
and in the realistic range (few GeV) can increase $Y_B$ by a factor
between $20\%$ (resonant case)  and $100\%$ (nonresonant case) compared to the $\Gamma_Y \to \infty$
limit.

\subsection{Phenomenology update}

\begin{figure}[!t]
\centering 
\epsfig{figure=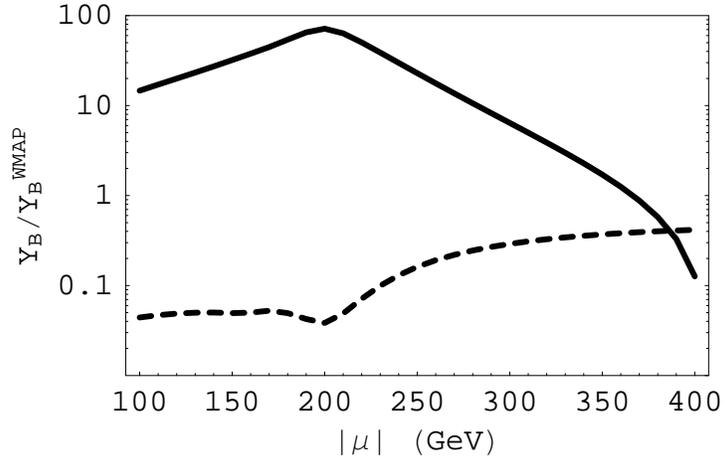,width=10cm}
\caption{$F_1$ (solid line) and $F_2$ (dashed line)  versus $|\mu|$   
with all other parameters fixed at the reference values of 
Table~\ref{tab:susy}. 
\label{fig:F1F2vsmu}
}
\end{figure}

In a consistent analysis $\Gamma_Y$ should not be
treated as an independent quantity (as we did in the last section for illustrative
purposes) but rather as a function of the MSSM parameters (as we did in
Section~\ref{sec:mssm}).  Doing so after numerically  solving the
transport equations, we  study the behavior of $F_{1,2}$
(Eq.~(\ref{eq:pheno1})) as a function of the MSSM parameters. For
illustration, we show in Fig.~\ref{fig:F1F2vsmu} the dependence of
$F_1$ (solid line) and $F_2$ (dashed line) on $|\mu|$, with all other 
input as in Table~\ref{tab:susy}. 
The plot highlights the resonant behavior of $F_1$ discussed in
\cite{Carena:1997gx,riotto,us}. The behavior of $F_2$ follows from the
fact that $F_2$ is proportional to $|\mu| (\Gamma_H+\Gamma_M^{-})^{-1/2}$: the dip at $|\mu|
\sim M_2 \sim 200$ GeV reflects the resonant enhancement of
$\Gamma_H$.
The overall scale of $F_{1,2}$ is set by $\Delta \beta $ which in turn
depends crucially on the CP-odd Higgs mass $m_A$~\cite{bubble}: here
we use $m_A = 150$ GeV but one should keep in mind that higher values
of $m_A$ can lead to sizable suppression of $F_{1,2}$.

Finally, we investigate the impact of electric dipole moment
(EDM) searches on this particular EWB scenario.  It has long been
recognized that, given the
spectrum of supersymmetric particles, constraints from the
electron~\cite{Regan:2002ta}, neutron~\cite{neutron}, and
nuclear~\cite{Romalis:2000mg} EDMs pose tight limits on the size of CP-violating phases (for a review see~\cite{Pospelov:2005pr}). 
These could ultimately enter in conflict with the
requirement of successful baryon asymmetry generation, making EDM
searches a great discriminating tool among theories of baryogenesis.


\begin{figure}[t]
\centering
\begin{picture}(300,250)  
\put(100,240){\framebox[1.0\width][c]{ At $m_{\tilde f} = $ 1 TeV }}
\put(-10,100){\makebox(50,50){\epsfig{figure=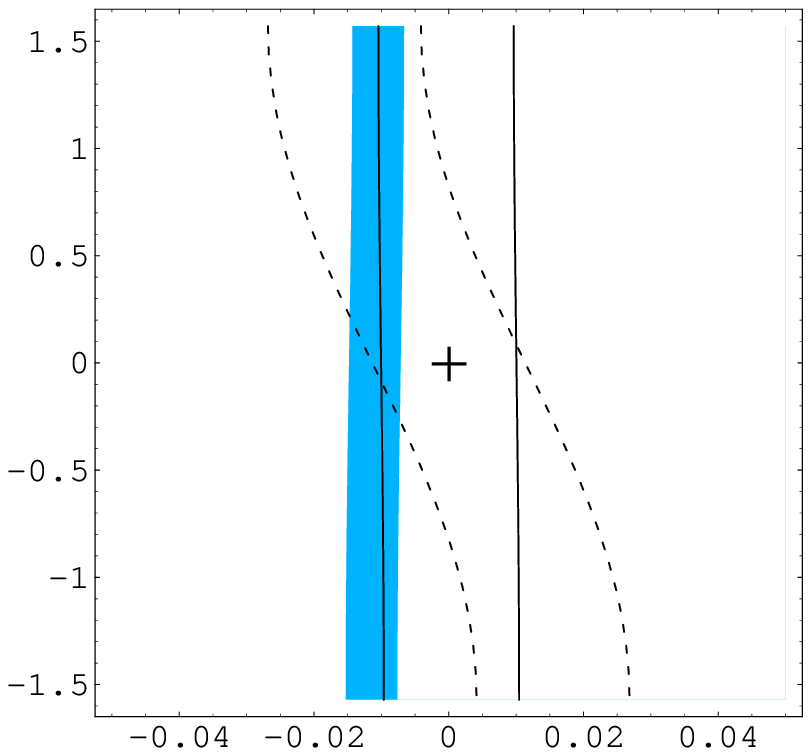,width=7.5cm}}}
\put(240,100){\makebox(50,50){\epsfig{figure=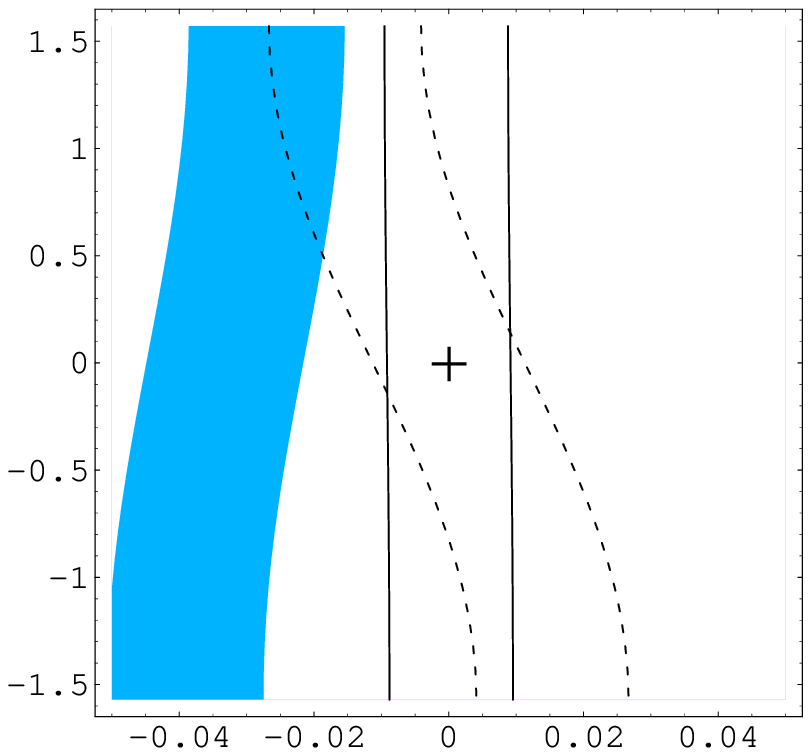,width=7.5cm}}}
\put(7,15){{\small $\phi_\mu$ (rad)}}
\put(260,15){{\small $\phi_\mu$ (rad)  }}
\put(-100,130){{\small $\phi_A$ }}
\put(-105,115){{\small (rad)}}
\put(150,130){{\small $\phi_A$ }}
\put(145,115){{\small (rad)}}
\put(280,200){{\small $d_e$ }}
\put(305,60){{\small $d_{n}$ }}
\put(205,75){{\scriptsize EWB}}
\put(-15,60){{\scriptsize EWB }}
\put(30,200){{\small $d_e$ }}
\put(55,60){{\small $d_{n}$ }}
\end{picture}

\caption{ Allowed bands in the $\phi_\mu$--$\phi_A$ plane implied by
consistency with the $95 \%$ C.L. limits on electron and neutron EDMs and baryogenesis.  The solid lines correspond to the constraint from the electron EDM ($|d_e| < 1.9
\times 10^{-27} e \cdot {\rm cm} $~\cite{Regan:2002ta}), and the dashed lines correspond to the neutron EDM
($|d_{n}| < 3.6 \times 10^{-26} e \cdot {\rm cm}$~\cite{neutron}).  These EDM constraints correspond to sfermion
masses ($m_{\tilde f}$) fixed at 1 TeV.  The shaded EWB band is the
region consistent with $Y_B$ from BBN~\cite{PDG} at 95$\%$ C.L.
(which includes the $Y_B$ range from WMAP~\cite{wmap}).
In the left-hand panel we use $|\mu| = M_2 = 200$ GeV (resonance peak),
while in the right-hand panel we use $M_2= 200$ GeV and $|\mu| = 250$ GeV
(off resonance).  The other supersymmetric masses are as specified in
the text.
\label{fig:bands}
}

\end{figure}

To illustrate this point  we plot in Fig.~\ref{fig:bands} the allowed
bands in the $\phi_\mu$--$\phi_A$ plane resulting from present limits on electron
and neutron EDMs and successful baryogenesis, for
a given choice of the SUSY mass parameters. Here, we have employed one-loop SUSY
contributions~\cite{Ibrahim:1997gj}.  We take the first- and
second-generation sfermions, as well as the gluinos, all degenerate at 1 TeV, while all other input
is fixed as in Table~\ref{tab:susy}.  In the left-hand panel we use
$M_2=|\mu| = 200$ GeV (resonance peak), while in the right we
use $M_2= 200$ GeV, $|\mu| = 250$ GeV.

\begin{figure}[!t]
\centering
\begin{picture}(300,175)  
\put(120,65){\makebox(50,50){\epsfig{figure=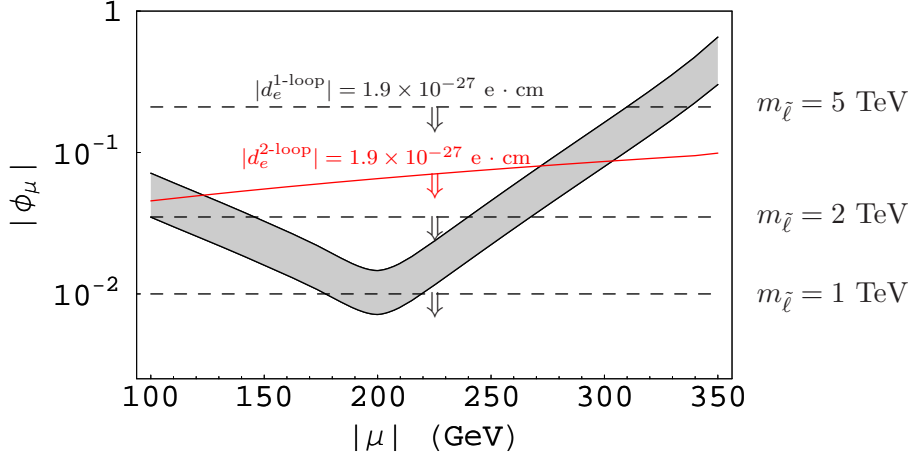,width=10cm}}}
\put(290,60){{\small $m_{\tilde{\ell}} = 1$ TeV}}
\put(290,90){{\small $m_{\tilde{\ell}} = 2$ TeV}}
\put(290,132){{\small $m_{\tilde{\ell}} = 5$ TeV}}
\put(100,138){{\scriptsize $\lvert d_e^{\text{1-loop}}\rvert = 1.9 \times 10^{-27}$ e $\cdot$ cm}}
\put(95,112){\color{red}\scriptsize $\lvert d_e^{\text{2-loop}}\rvert = 1.9 \times 10^{-27}$ e $\cdot$ cm}
\put(165,126){$\Downarrow$}
\put(165,101){\color{red}$\Downarrow$}
\put(165,85){$\Downarrow$}
\put(165,56){$\Downarrow$}
\end{picture}
\caption{ We plot in the $|\phi_\mu |$-$|\mu|$ plane the region
consistent with EWB (gray shaded band), the $|d_e^\text{1-loop}| = 1.9 \times
10^{-27} \, {\rm e} \cdot \, {\rm cm} $ (95 $\%$ C.L. limit) curves
for various values of the first-generation slepton masses (dashed
horizontal lines), and the $|d_e^\text{2-loop}| = 1.9 \times
10^{-27} \, {\rm e} \cdot \, {\rm cm} $ curve.  For a given choice of mass parameters, the allowed parameter region lies below the EDM curves.  The baryon-to-entropy ratio is required to be
in the range $ 4.8 \times 10^{-11} < Y_B < 9.8 \times
10^{-11} $~\cite{PDG},  and the SUSY parameters are as in Table~\ref{tab:susy}. 
\label{fig:phimu}
}
\end{figure}

Figure~\ref{fig:bands} illustrates the complementarity of various EDM
measurements in constraining the new CP-violating phases in general. It also
shows that in this particular scenario it is the electron EDM that
poses the strongest constraints on electroweak baryogenesis.  
 In order to quantify the dependence of the EDM constraints on the heavy sfermion masses,  we plot in Fig.~\ref{fig:phimu} the region in the $|\phi_\mu
|$-$|\mu|$ plane that is consistent with EWB (gray shaded band) along
with the $|d_e^\text{1-loop}| = 1.9 \times 10^{-27} \, {\rm e} \cdot \, {\rm cm} $
(95 $\%$ C.L. limit) curves for various values of the first generation
slepton masses (assumed degenerate).  For a given slepton mass, the
region in the $|\phi_\mu|$-$|\mu|$ plane consistent with EDM
constraints lies {\em below} the dashed line. In the same figure, we also plot the $\abs{d_e^{\text{2-loop}}} =1.9\times 10^{-27}\,\text{e}\cdot\text{cm}$ curve (solid red line) from two-loop SUSY contributions \cite{2-loop}.
Several key features emerge from Figs.~\ref{fig:bands} and
\ref{fig:phimu}:
\begin{itemize}
\item[(i)] In the range of $\mu$ and $M_2$ we are considering, $\abs{d_e}$ is dominated by the one-loop contributions for slepton masses below 1--2 TeV, while the two-loop effects become dominant for slepton masses larger than 2--3 TeV.
\item[(ii)] In the case of resonant EWB, which requires the smallest
amount of CP violation, the electron EDM constraint requires
slepton masses to be heavier than 1 TeV.  
\item[(iii)] Two-loop contributions to $d_e$ imply that EWB cannot occur too far off resonance (see Fig.~\ref{fig:phimu}), even in the limit of very heavy sleptons.
\end{itemize} 
Additional constraints on higgsino-mediated electroweak baryogenesis 
do arise from the phenomenology of indirect dark matter detection 
in the MSSM, and they are investigated in Ref.~\cite{profumoetal}. 

Before concluding, we emphasize that the constraints implied by Figs.~\ref{fig:bands} and
\ref{fig:phimu} are specific to the MSSM, and that the extensions of the MSSM discussed in Ref.~\cite{nonminimal} and elsewhere can lead to different phenomenological conclusions. In particular, extended Higgs sector models with additional scalar degrees of freedom can give rise to a strong, first-order electroweak phase transition without requiring a light ${\tilde t}_R$. In this case, resonances in the stop sector may enhance the importance of CP-violation associated with the tri-scalar terms ({\em e.g.}, $\phi_A$), and the information provided by the neutron and neutral atom EDM searches would become more important than for the MSSM scenario considered here. In addition, we also note that there could exist additional, $\mathcal{O}(1)$ corrections to $Y_B$ associated with computations of the sphaleron rates, bubble profile, and Majorana gaugino transport that we have not addressed here.

\section{Conclusions}
\label{sec:conclusions}

The present work is part of a broader program initiated in~\cite{us}
whose goal is to systematically reduce uncertainties in EWB
calculations induced by transport phenomena.  The main new results of
this work are:

\begin{itemize}

\item We have calculated the contribution to the quantum Boltzmann
equations due to decays and inverse decays induced by tri-scalar and
Yukawa-type interactions.  We have performed the calculation in the
Closed Time Path formalism to leading non-trivial
order in the ratios
$\varepsilon_\mu= \mu/T$, $\varepsilon_p = \Gamma/\omega$,
$\varepsilon_d = v_w k_{\rm eff}/\omega$.

\item Specializing to the case of MSSM, we have derived the (inverse)
decay rate due to top-quark Yukawa interactions, their supersymmetric
tri-scalar counterparts, and the soft SUSY-breaking tri-scalar
interactions proportional to $y_t$.  These rates are of ${\cal
O}(\alpha_s^0)$, and -- when not vanishing due to threshold effects -- they
can be of comparable size or larger than the ${\cal O}(\alpha_s)$
contributions from scattering processes.

\item We have revisited the fast-$\Gamma_Y$
approximation~\cite{HN}, which consists in taking the rate $\Gamma_Y$
of Yukawa-induced processes as large compared to all other relevant
time-scales. We have found previously-unnoticed corrections to the baryon density that enter at linear order in the $1/\Gamma_Y$-expansion, whose inclusion shows that this expansion in fact breaks down. The approximation is sound in
the unbroken phase, where Yukawa processes are, indeed, fast on the scale of diffusion processes. But in the broken phase, the rates
$\Gamma_M$, $\Gamma_H$ associated with relaxation processes
can be as large as or larger than $\Gamma_Y$, even after including the
${\cal O}(\alpha_s^0)$ contributions to $\Gamma_Y$.  The interplay of
these competing processes is quite significant, and a quantitative analysis requires performing a numerical solution to the transport equations for realistic, finite values of $\Gamma_Y$. 
For the parameter choices we considered, keeping $\Gamma_Y$
finite can increase $Y_B$ by a factor between $20\%$ and $100\%$
compared to the $\Gamma_Y \to \infty$ limit.

\item We have updated our previous~\cite{us} analysis of the
connection between EDM constraints  and EWB. Even within present uncertainties, the
simultaneous requirement of successful EWB and consistency with EDM
upper limits, poses stringent constraints on the size of SUSY CP
violating phases and mass spectrum.  For example, for any value of the
CP violating phases, successful baryogenesis and one-loop EDM constraints force
the slepton masses to be heavier than $\sim 1$ TeV.   Bounds of this type
will be sharpened by future EDM experiments and can be tested at
future collider experiments.

\end{itemize}

\acknowledgments 
We wish to thank C. Wagner,
M. Carena, and M. Wise for useful discussions, and A. Pilaftsis for pointing us to the work in \cite{2-loop} on two-loop SUSY contributions to EDMs. CL is grateful to the high energy and nuclear theory groups at Caltech for their hospitality during portions of this work. 
The work of MJRM and ST was supported by 
U.S. Department of Energy contract DE-FG02-05ER41361 and by a National  
Science Foundation Grant PHY00-71856. VC was supported by Caltech 
through the Sherman Fairchild fund.  The work of CL was supported by the U.S. Department of Energy contract DE-FG02-00ER41132.

\appendix

\section{ $\mathcal{I}_B$ and $\mathcal{I}_F$ in terms of 
one-dimensional integrals}
\label{app:app1}

Performing the $\omega_L$ integral in Eq.~(\ref{eq:IB}) and the
$\omega_\phi$ integral in Eq.~(\ref{eq:IF}) yields
\begin{equation}
\begin{split}
& \mathcal{I}_B (A_s \, ; \,   m_R,m_L,m_H) =  \frac{\abs{A_s}^2}{16\pi^3} \,
\int_{m_R}^\infty d\omega_R  \, h_B(\omega_R)  \\
& \ \ \ \times 
\biggl\{\log\left(\frac{e^{\omega_R/T} - 
e^{\omega_L^+/T}}{e^{\omega_R/T} - e^{\omega_L^-/T}}
\frac{e^{\omega_L^-/T}-1}{e^{\omega_L^+/T}-1}\right)
\bigl[\theta(m_R-m_L-m_H) - \theta(m_L-m_R-m_H)\bigr] \\
&\qquad\qquad\qquad\qquad\quad + \log\left(\frac{e^{-\omega_R/T} - e^{\omega_L^+/T}}{e^{-\omega_R/T} - e^{\omega_L^-/T}}\frac{e^{\omega_L^-/T}-1}{e^{\omega_L^+/T}-1}\right)\theta(m_H-m_R-m_L)\biggr\}
\end{split}
\end{equation} 
and 
\begin{equation}
\begin{split}
& \mathcal{I}_F (m_1,m_2,m_\phi)  = -  \frac{1}{16\pi^3} \,
\left(m_1^2 + m_2^2 - m_\phi^2 \right) \, 
\int_{m_1}^\infty d\omega_1 h_F(\omega_1) \\
&\ \ \ \times 
\biggl\{\log\left(\frac{e^{\omega_\phi^+/T} - 1}{e^{\omega_\phi^-/T} - 1}
\frac{e^{\omega_\phi^-/T} + e^{\omega_1/T}}{e^{\omega_\phi^+/T} + 
e^{\omega_1/T}}\right)[\theta(m_1 - m_2 - m_\phi) - 
\theta(m_\phi - m_1 - m_2)\bigr] \\
& \qquad\qquad\qquad\quad+ \log\left(\frac{e^{\omega_\phi^+/T} - 
1}{e^{\omega_\phi^-/T} - 1}\frac{e^{\omega_\phi^-/T} + 
e^{-\omega_1/T}}{e^{\omega_\phi^+/T} + e^{-\omega_1/T}}\right)
\theta(m_2 - m_1 - m_\phi)\biggr\}
\end{split}
\end{equation}
where 
\begin{equation}
h_{F,B} (x)  = - \frac{e^{x/T}}{(e^{x/T} \pm 1)^2}  \ . 
\end{equation}

\section{Details of the source derivation } 
\label{app:app2}

In this Appendix we give some details of the derivation of the 
source terms reported in Section~\ref{sec:mssm}. 
We first relate the scalar Higgs chemical potentials $\mu_{H_{u,d}}$
to the Higgsino chemical potential $\mu_{\tilde H}$ and then show
how to further simplify the final expression by use of mass relations
among weak doublet partners.

Recall that the Higgsino chemical potential $\mu_{\tilde H}$
corresponds to the vector charges, $n_{\tilde H^{+,0}} =
\bar\Psi_{\tilde H^{+,0}}\gamma^0\Psi_{\tilde H^{+,0}}$, for
the Dirac fields introduced in Eq.~(\ref{eq:DiracPsiH}), which combine
$u$- and $d$-type Higgsino densities. To determine the scalar Higgs
density that is kept in equilibrium with the Higgsino vector charge
density {\em via} gaugino interactions, we examine their interactions
in the MSSM Lagrangian, written in terms of the Dirac fields
$\Psi_{\tilde H^{+,0}}$, and the four-component gaugino fields:
\begin{equation}
\Psi_{\tilde W^+} = \begin{pmatrix} \tilde W^+ \\ \tilde W^{-\dag} \end{pmatrix},
\qquad
\Psi_{\tilde W^0} = \begin{pmatrix} \tilde W_3 \\ \tilde W_3^\dag \end{pmatrix},
\qquad
\Psi_{\tilde B} = \begin{pmatrix} \tilde B \\ \tilde B^\dag \end{pmatrix}.
\end{equation}
The charged wino field $\Psi_{\tilde W^+}$ is a Dirac spinor, for
which a vector charge density can also be defined, while the neutral
fields $\Psi_{\tilde W^0,\tilde B}$ are Majorana spinors,
whose vector charge density is zero. In terms of these fields, the
Higgs-Higgsino-gaugino interactions are:
\begin{equation}
\label{eq:Lgaugino}
\begin{split}
\mathcal{L}_{H\tilde H\tilde V} = &-\frac{g_1}{\sqrt{2}}\Bigl[ \bar\Psi_{\tilde H^+}(H_d^{-*} P_L + e^{i\phi_\mu}H_u^{+}P_R )\Psi_{\tilde B} + \bar\Psi_{\tilde H^0}(H_d^{0*}P_L - e^{i\phi_\mu}H_u^0 P_R)\Psi_{\tilde B}\Bigr] \\
&- \frac{g_2}{\sqrt{2}}\Bigl[\bar\Psi_{\tilde H^+}(-H_d^{-*}P_L + e^{i\phi_\mu}H_u^+ P_R)\Psi_{\tilde W^0} + \bar\Psi_{\tilde H^0}(H_d^{0*}P_L + e^{i\phi_\mu}H_u^0 P_R)\Psi_{\tilde W^0}\Bigr] \\
&- g_2\Bigl[\bar\Psi_{\tilde H^+}(H_d^{0*}P_L + e^{i\phi_\mu}H_u^0 P_R)\Psi_{\tilde W^+} + \bar\Psi_{\tilde W^+}(H_d^{-*}P_L - e^{i\phi_\mu}H_u^+ P_R)\Psi_{\tilde H^0}^C \Bigr] \\
&+ \text{h.c.}
\end{split}
\end{equation}

The combinations of scalar fields appearing in each term of
Eq.~(\ref{eq:Lgaugino}) tell us which densities are kept in
equilibrium with the Higgsino densities by fast gaugino
interactions. To illustrate, consider the second term on the RHS that
couples the ${\tilde H}_0$ and ${\tilde B}_0$ fields to the
neutral Higgs fields. Using $\mu_{\tilde B}=0$ we see from
Eqs. (\ref{eq:lagYf},\ref{eq:sourceF}) that this term in
$\mathcal{L}_{H\tilde H\tilde V}$ will generate source terms
for ${\tilde H}_0$ given by
\begin{eqnarray}
S_{\tilde H^0} &=& -\frac{g_1^2}{2}\left[(\mu_{{\tilde H}^0}-\mu_{H_u^0})\mathcal{I}_F(m_{\tilde H^0},m_{\tilde B}, m_{H_u^0})+(\mu_{{\tilde H}^0}+\mu_{H_d^0})\mathcal{I}_F(m_{\tilde H^0},m_{\tilde B}, m_{H_d^0})\right] \\
&=& \quad \frac{g_1^2}{2}\left[ (\mu_{H^0}-\mu_{{\tilde H}^0}) \mathcal{I}_{F,\, H}^{(+)} + \mu_{h^0} \mathcal{I}_{F,\, H}^{(-)}\right] \ \ \ ,
\end{eqnarray}
where
\begin{eqnarray}
\mu_{H^0} & \equiv & = \frac{1}{2}\left(\mu_{H_u^0}-\mu_{H_d^0}\right) \\
\mu_{h^0} & \equiv & = \frac{1}{2}\left(\mu_{H_u^0}+\mu_{H_d^0}\right) \\
\mathcal{I}_{F,\, H}^{(\pm)} & = & \mathcal{I}_F(m_{\tilde H^0},m_{\tilde B}, m_{H_u^0}) \pm
\mathcal{I}_F(m_{\tilde H^0},m_{\tilde B}, m_{H_d^0})\ \ \ .
\end{eqnarray}
Similar expressions follow from the other terms in
Eq. (\ref{eq:Lgaugino}) (assuming the ${\tilde W}$ densities
vanish). 
The assumption of
\lq\lq fast" supergauge interactions then leads to:
\footnote{This is tantamount to assuming that 
$g_1^2 \mathcal{I}_{F,\, H}^{(+)}$ is sufficiently large compared to
the other transport coefficients so that $\mu_{H^0}-\mu_{{\tilde    
H}^0}\approx 0$. 
We leave for future work an explicit test of this assumption. 
A comprehensive analysis that allows for $\mu_{H^0}\not= \mu_{{\tilde
H}^0}$ should also include the effects of non-vanishing gaugino
densities, since gauginos play an essential role in this departure
from chemical equilibrium. Since the neutral gauginos are Majorana
fermions and possess no vector current density, such an analysis will
require in turn a study of the axial vector analog of
Eq. (\ref{eq:fermion1}) \cite{us}, a task that goes beyond the scope
of the present work.} 
\begin{eqnarray}
\label{eq:vectorscalarrelation}
\mu_H & \equiv & \frac{1}{2}\left(\mu_{H_u} - \mu_{H_d}\right) = \mu_{\tilde H} \\
\mu_{h} &\equiv& \frac{1}{2}\left(\mu_{H_u} + \mu_{H_d}\right)
\end{eqnarray}
and 
\begin{equation}
\label{eq:HuHdHh}
\mu_{H_u} = \mu_H + \mu_h\qquad \mu_{H_d} = \mu_h - \mu_H \ \ \ ,
\end{equation}
where $\mu_H$ and $\mu_h$ refer to the common chemical potentials for
the charged and neutral Higgs scalars. 
Adding together the top and stop sources in
Eq.~(\ref{eq:tRsource},\ref{eq:stRsource}) and using the relations
(\ref{eq:equalmus},\ref{eq:vectorscalarrelation},\ref{eq:HuHdHh})
gives for the Yukawa source for the density $T$:
\begin{equation}
\label{eq:Tsource}
\begin{split}
S_T^Y = -N_C y_t^2\biggl\{(\mu_T - \mu_Q - \mu_H)  \Bigl[&\mathcal{I}_B(A_t ; m_{\tilde t_R},m_{\tilde t_L},m_{H_u^0}) + \mathcal{I}_B(A_t ; m_{\tilde t_R},m_{\tilde b_L},m_{H_u^+}) \\
+&\mathcal{I}_B(\mu ; m_{\tilde t_R},m_{\tilde t_L},m_{H_d^0})  + \mathcal{I}_B(\mu ; m_{\tilde t_R},m_{\tilde b_L},m_{H_d^-}) \\
+ &\mathcal{I}_F(m_{\tilde H^+},m_{t_L},m_{\tilde t_R}) + \mathcal{I}_F(m_{\tilde H^0},m_{b_L},m_{\tilde t_R}) \\
+&\mathcal{I}_F(m_{t_R},m_{t_L},m_{H_u^0}) + \mathcal{I}_F(m_{t_R},m_{b_L},m_{H_u^+}) \\
+& \mathcal{I}_F(m_{t_R},m_{\tilde H^0},m_{\tilde t_L}) + \mathcal{I}_F(m_{t_R},m_{\tilde H^+},m_{\tilde b_L}) \Bigr] \\
+ \mu_h\Bigl[ &\mathcal{I}_B(\mu; m_{\tilde t_R},m_{\tilde t_L},m_{H_d^0}) + \mathcal{I}_B(\mu ; m_{\tilde t_R},m_{\tilde t_L},m_{H_d^-}) \\ 
- &\mathcal{I}_B(A_t ; m_{\tilde t_R},m_{\tilde t_L}, m_{H_u^0}) - \mathcal{I}_B(A_t ; m_{\tilde t_R}, m_{\tilde b_L} , m_{H_u^+}) \\
- &\mathcal{I}_F(m_{t_R},m_{t_L},m_{H_u^0}) - \mathcal{I}_F(m_{t_R},m_{b_L},m_{H_u^+})\Bigr]\biggr\}
\end{split}
\end{equation}
We can simplify further by noting that the masses of weak doublet
partners are the same:
\begin{subequations}
\label{eq:massrelations}
\begin{align}
m_{t_L} &= m_{b_L} \equiv m_Q \\
m_{\tilde t_L} &= m_{\tilde b_L} \equiv m_{\tilde Q} \\
m_{H_u^+} &= m_{H_u^0} \equiv m_{H_u} \\
m_{H_d^-} &= m_{H_d^0} \equiv m_{H_d} \\
m_{\tilde H^+} &= m_{\tilde H^0} = \abs{\mu} ~ . 
\end{align}
\end{subequations}
With the notation for the masses introduced here we arrive 
at our final result of Eq.~(\ref{eq:GammaY}).

\section{Analytic corrections of  $\mathcal{O}(1/\Gamma_{Y,ss})$}
\label{app:app3}

In this Appendix we solve the transport equations in powers of
$1/\Gamma_{Y,ss}$ and show that the analytic solutions obtained in the
$\Gamma_{Y} \to \infty$ limit can receive $\mathcal{O}(1)$ corrections for
realistic choices of all the competing rates ($\Gamma_{Y,H,M}$).

The zeroth-order solution in $1/\Gamma_{Y,ss}$ is obtained by
considering the combination of Eqs.~(\ref{eq:QTEs}) that is
independent of $\Gamma_Y$ and $\Gamma_{ss}$.  Letting $D_h$ and $D_q$
be the diffusion constants for Higgs and quark superfields,
respectively [see Eq.~(\ref{eq:nrelation})], letting the densities be
a function of ${\bar z}=\abs{\vect{x}+\vect{v}_w t}$ (the
co-moving distance from the bubble surface along its normal), and
neglecting small corrections proportional to $\Gamma_M^+$ for
simplicity, we obtain:
\begin{equation}
\label{eq:solve1}
\begin{split}
D_q[2T''(\bar{z}) + Q''(\bar{z})] + D_h H''(\bar{z}) & 
- v_w[2T'(\bar{z}) + Q'(\bar{z}) + H'(\bar{z})]  
 \\ = &
 \Gamma_M^{-}(\bar{z}) \biggl[\frac{T(\bar{z})}{k_T} 
- \frac{Q(\bar{z})}{k_Q}\biggr] 
+ \Gamma_h(\bar{z}) \frac{H(\bar{z})}{k_H} 
- S_{TOT}^{\CPV} (\bar{z})
\end{split}
\end{equation}
where $S_{TOT}^{\CPV}=S_{\tilde H}^{\CPV}+S_{\tilde t}^{\CPV}$ and
$f^\prime = \partial f/\partial{\bar z}$.  
The approximate chemical equilibrium enforced by Yukawa and 
strong sphaleron processes implies that the combinations 
\begin{eqnarray}
\label{eq:solve2}
\delta_Y & = & \frac{T}{k_T} - \frac{Q}{k_Q} - \frac{H}{k_H}\\
\nonumber \delta_{ss} & = & -\left(\frac{2Q}{k_Q} - \frac{T}{k_T} +
\frac{9(Q+T)}{k_B}\right)
\end{eqnarray}
tend to zero in the limit $\Gamma_{Y,ss} \to \infty$, so that we can
formally expand in $1/\Gamma_{Y,ss}$ and treat for bookkeeping
purposes $\delta_Y\sim 1/\Gamma_Y$ and $\delta_{ss}\sim
1/\Gamma_{ss}$.  The relations between the $Q$, $T$, and $H$ densities,  up
to order $1/\Gamma_{Y,ss}$ are then:
\begin{eqnarray}
\label{eq:solve3}
Q & = &  \frac{k_Q(k_B-9k_T)}{k_H(9k_T+9k_Q+k_B)} \, 
\left(H+k_H \delta_Y\right) - \frac{k_B k_Q}{(9k_T+9k_Q+k_B)}
\, \delta_{ss}\\
\nonumber
T & = & 
\frac{k_T(2k_B+9k_Q)}{k_H(9k_T+9k_Q+k_B)}
\, \left(H+k_H \delta_Y\right) 
-\frac{k_B k_T}{(9k_T+9k_Q+k_B)} 
\, \delta_{ss}~.
\end{eqnarray}
Substituting these expressions back into Eq.~(\ref{eq:solve1}), we obtain the equation
for $H$: 
\begin{equation}
\label{eq:solve5}
v_w H^\prime (\bar{z})  -{\bar D} H^{\prime\prime} (\bar{z})  
= -{\bar \Gamma} H (\bar{z}) +
{\bar S} (\bar{z}) 
+\delta{\bar S} (\bar{z}) 
\end{equation}
where 
\begin{eqnarray}
\nonumber
{\bar D} & = & \frac{D_h\Delta+D_q K}{(K+\Delta)}\\
\nonumber
{\bar \Gamma} & = & \frac{\Delta}{k_H(K+\Delta)}\left(\Gamma_M^-+\Gamma_H\right) \\
\label{eq:solve6}
{\bar S} 
& = & \frac{\Delta}{(K+\Delta)} S_{TOT}^{\CPV}\\
\nonumber
K & = & 9k_Tk_Q + k_B k_Q + 4 k_T k_B\\
\nonumber
\Delta & = & k_H\left( 9 k_T + 9 k_Q +k_B\right)\  \  \ ,
\end{eqnarray}
and 
\begin{equation}
\label{eq:solve7}
\delta{\bar S} = \frac{k_H}{(K+\Delta)}
\biggl[ k_B(2k_T+k_Q)(v_w \delta_{ss}^\prime -D_q\delta_{ss}^{\prime\prime}) 
- K (v_w \delta_{Y}^\prime -D_q\delta_{Y}^{\prime\prime}) 
- (\Delta/k_H)\Gamma_M^-\delta_Y  \biggr]
\end{equation}
represents a correction to the effective source $\bar S$ for the Higgs
density $H$.  The functions $\delta_Y$ and $\delta_{ss}$ appearing in
Eq.~(\ref{eq:solve7}) are determined by substituting the lowest order
solution $H_0$ into Eqs.~(\ref{eq:QTEs}) and read:
\begin{subequations}
\label{deltaYss}
\begin{align}
\delta_Y(\bar{z}) &= -\frac{1}{\Gamma_Y} \biggl[D_h H_0''(\bar{z}) -
v_w H_0'(\bar{z}) - \Gamma_H(\bar{z})\frac{H_0(\bar{z})}{k_H} +
S_{\tilde h}(\bar{z})\biggr] \label{deltaY} \\ 
\delta_{ss}(\bar{z}) &=
-\frac{1}{\Gamma_{ss}}\frac{k_B}{k_H}\frac{k_Q + 2k_T}{k_B + 9k_T
+9k_Q}\bigl[D_q H_0''(\bar{z}) - v_w
H_0'(\bar{z})\bigr]. \label{deltass}
\end{align}
\end{subequations}
Although in the unbroken phase $\delta_{Y,ss} \sim
\Gamma_{\rm diff}/\Gamma_{Y,ss} \times H_0 \ll H_0$, in the broken phase,
they can be sizable, with $\delta_Y \gg \delta_{ss}$. 

All previous treatments have neglected the $\delta\bar S$ term in
Eq.~(\ref{eq:solve5}) and thus find only the leading-order solution for
$H$. Then the only $1/\Gamma_{Y,ss}$ effects appear to be the
$\delta_{Y,ss}$ terms in Eqs.~(\ref{eq:solve3}). However,
$\delta\bar S$ induces $\mathcal{O}(1/\Gamma_{Y,ss})$ corrections to
the density $H$ obtained by solving Eq.~(\ref{eq:solve5}), which must
be substituted back into Eqs.~(\ref{eq:solve3}) to give the full $Q,T$
densities to order $1/\Gamma_{Y,ss}$.
Using the simplified bubble wall profile as in Ref.~\cite{us} (with
constant sources in the region $0<{\bar z}< L_w$), the explicit
solution to Eq.~(\ref{eq:solve5}) in the region of unbroken
electroweak symmetry (${\bar z} <0$), that drives the weak sphaleron
processes,  reads
\begin{equation}
\label{eq:solve8}
 H_{<}({\bar z}) = \left[ \frac{1}{{\bar
 D}\kappa_{+}} \, \int_0^\infty dy\, e^{\rm -\kappa_{+} y}\,
 \left({\bar S}(y)
+\delta{\bar S}(y)\right)
+ \frac{\delta{\bar S}_<(0)}{v_w}\left(\frac{1}{\kappa_{+}}-{\bar
 z}\right) \right] e^{v_w{\bar z}/{\bar D}}\ \ \ ,
\end{equation}
with
\begin{equation}
\kappa_\pm = \frac{1}{2}\left( v_w\pm \sqrt{v_w^2+
4{\bar\Gamma}{\bar D}}\right) \ ,
\end{equation}
and $\delta\bar S_<(0)$ is the value that $\delta \bar S$ takes at $\bar z=0$ approaching from the left.
The ${\cal O}(1/\Gamma_{Y,ss})$ contributions to $H$ live in the terms
containing $\delta{\bar S}$. The largest effect arises from the
presence of $\delta{\bar S}(y)$ inside the integral.  The overall size
of $\delta \bar{S}$ is dominated by the term in Eq.~(\ref{eq:solve7})
proportional to $\delta_Y \Gamma_M^-$. Moreover, the typical size of
$\delta_Y$ is set by $(\Gamma_H/\Gamma_Y) \cdot (H_0/k_H)$, leading to:
\begin{equation}
\label{eq:solve9}
\frac{\delta{\bar S}}{{\bar S} }\sim
\left(\frac{\Gamma_H}{\Gamma_Y}\right)
\frac{\sqrt{r_\Gamma} \, \Gamma_M^- \, L_w}{\sqrt{\bar{D} (\Gamma_M^- + 
\Gamma_H)}} 
\end{equation}
with $r_\Gamma = \Delta/[k_H(K+\Delta)] \sim 0.07$~.  Using earlier
estimates of $\Gamma_H$ and $\Gamma_M^-$~\cite{HN}, we find $\delta
\bar{S}/\bar{S} \sim 0.1$, indeed a small correction.  However, when
using $\Gamma_H$, $\Gamma_M^-$ and $\Gamma_Y$ as calculated in
Ref.~\cite{us} and the present work within the CTP framework, we find  
$\delta{\bar S}/{\bar S} \sim 1$, thus invalidating the 
assumption of fast $\Gamma_Y$ rates. 

In conclusion, the large $\delta_{Y,ss}$ corrections in the broken
phase induce large corrections to the effective source for the Higgs
density, which in turn induce large corrections to $Q,T$
themselves. What past treatments have derived  correctly are the
$1/\Gamma_{Y,ss}$ corrections to the \emph{relation} between $Q,T$, and
$H$ (that is, Eq.~(\ref{eq:solve3})), but not the corrections to $H$ itself. Yet this correction, it
turns out, is the biggest piece of all.

\end{document}